\documentclass[nofootinbib]{revtex4}

\usepackage{graphicx}        
\usepackage{bm}
\usepackage{amsmath,amssymb,amsfonts} 
\usepackage{natbib}

\newcommand{\der}[2]{\frac{\partial #1}{\partial #2}}

\def\Mdot{\dot M}
\def\msun{\rm M_{\odot}}
\newcommand{\bS}{\mbox{\boldmath$S$}}

\begin{document}

\title{Black hole accretion discs\footnote{Chapter 1 in \textsl{ Astrophysics of Black Holes -- From fundamental aspects to latest developments}, Ed. Cosimo Bambi, Springer: Astrophysics and Space Science Library 440, (2016); DOI: 10.1007/978-3-662-52859-4.}}
\author{Jean-Pierre Lasota\footnote{{lasota@iap.fr}}}
\affiliation{
Institut d'Astrophysique de Paris, CNRS et Sorbonne Universit\'es, UPMC Universit\'e Paris~06, UMR 7095, 98bis Boulevard Arago, 75014 Paris, France}
\affiliation{Nicolaus Copernicus Astronomical Center, Bartycka 18, 00-716 Warsaw, Poland \\ 
}
%
%


\begin{abstract}This is an introduction to models of accretion discs around black holes. After a presentation of the non--relativistic
equations  describing the structure and evolution of geometrically thin accretion discs we discuss their steady--state solutions and
compare them to observation. Next we describe in detail the thermal--viscous disc instability model and its application to dwarf novae
for which it was designed and its X--ray irradiated--disc version which explains the soft X--ray transients, i.e. outbursting black--hole low--mass X--ray binaries.
We then turn to the role of advection in accretion flows onto black holes illustrating its action and importance with a toy model describing both
ADAFs and slim discs. We conclude with a presentation of the general-relativistic formalism describing accretion discs in the Kerr space-time.
\end{abstract}
\maketitle
\section{Introduction}
\label{sec:intro}

The author of this chapter is old enough to remember the days when even serious 
astronomers doubted the existence of accretion discs and scientists snorted
with contempt at the suggestion that there might be such things as black holes; 
the very possibility of their existence was rejected, and the idea of black holes
was dismissed  as a fancy of eccentric theorists. Today, some 50 years later, 
there is no doubt about the existence of accretion discs and black holes; both 
have been observed and shown to be ubiquitous in the Universe. The spectacular 
ALMA image of the protostellar disc in HL Tau \cite{ALMAHLtau} is  breathtaking 
and we can soon expect to see the silhouette of a supermassive black hole in 
the center of the Galaxy in near infrared \cite{GRAVITY} or millimeter-waves \cite{EHT}.

Understanding accretion discs around black holes is interesting in itself because of the fascinating and complex physics
involved but is also fundamental for understanding the coupled evolution of galaxies and their nuclear black holes, i.e.
fundamental for the understanding the growth of structures in the Universe. The chance that inflows onto black holes
are strictly radial, as assumed in many models, are slim.

The aim of the present chapter is to introduce the reader to models of accretion discs around black holes.
Because of the smallness of black holes the sizes of their accretion discs span several orders of magnitude: from
close to the horizon up 100 000 or even 1 000 000 black-hole radii. This implies, for example, that the temperature
in a disc around a stellar--mass black hole varies from $10^7$\,K, near the its surface, to $\sim 10^3$\,K near the disc's
outer edge at $10^5$ black-hole radii, say. Thus studying black hole accretion discs allow the study of physical regimes
relevant also in a different context and inversely, the knowledge of accretion disc physics in other systems such as e.g. 
protostellar discs or cataclysmic variable stars, is useful or even necessary for understanding the discs around black holes.

Section \ref{sec:viscosity} contains a short discussion of the disc driving mechanisms and introduces the $\alpha$--prescription
used in this chapter. In Section \ref{sec:thin} after presenting the general framework of the geometrically thin disc model we
discuss the properties of stationary solutions and the Shakura--Sunyaev solution in particular. The dwarf-nova disc instability
model and its application to black-hole transient sources is the subject of Section \ref{sec:DI}. The role of advection in accretion
onto black holes is presented in Section \ref{sect:advection} with the main stress put on high accretion rate flows. Finally, Section \ref{sec:kerr} and \ref{sec:accretion_Kerr} about the
general--relativistic version of the accretion disc equations concludes the present chapter.

\textsc{Notations and definitions}

The Schwarzschild radius (radius of a non-rotating black hole) is
\begin{equation}
R_S=\frac{2GM}{c^2}= 2.95\times 10^{5} \frac{M}{\rm M_{\odot}}\, \rm cm,
\end{equation}
where $M$ is the mass of the gravitating body and $c$ the speed of light.

The Eddington accretion rate is defined as
\begin{equation}
\dot{M}_{\rm Edd}=\frac{L_{\rm Edd}}{\eta c^2}=\frac{1}{\eta}\frac{4\pi GM}{c\kappa_{es}}=\frac{1}{\eta}\frac{2\pi c R_S}{\kappa_{es}}=1.6\times 10^{18}\eta_{0.1}^{-1} \frac{M}{\rm M_{\odot}}\,\rm g\,s^{-1},
\end{equation}
where $\eta=0.1\eta_{0.1}$ is the radiative efficiency of accretion, $\kappa_{es}$ the electron scattering (Thomson) opacity.
  
We will often use accretion rate measured in units of Eddington accretion rate:
\begin{equation}
\label{eq:mdotEdd}
\dot m=\frac{\dot M}{\dot{M}_{\rm Edd}}.
\end{equation}
\\

\textsc{Additional reading:} There are excellent general reviews of accretion disc physics, they can be found in references \cite{Blaes14}, \cite{FKR} \cite{Katoetalbook} and \cite{Spruit10}.
\section{Disc driving mechanism; viscosity}
\label{sec:viscosity}

In recent years there have been an impressive progress in understanding
the physical mechanisms that drive disc accretion. It is now obvious that the
turbulence in ionized Keplerian discs is due to the Magneto-Rotational Instability
(MRI) also known as the Balbus-Hawley mechanisms \cite{BalbusHawley91},\cite{Balbus11}. 
However, despite these developments, numerical simulations, even in their global, 3D form suffer still
from weaknesses that make their direct application to real accretion flows almost infeasible. 

One of the most serious problems is the value of the ratio of the (vertically averaged) total stress to thermal (vertically averaged) pressure
\begin{equation}
\label{eq:numeralpha}
\alpha=\frac{\langle{\tau_{r\varphi}}\rangle_z}{\langle P \rangle_z}
\end{equation}
which according to most MRI simulation is $\sim 10^{-3}$ whereas observations of dwarf nova
decay from outburst unambiguously show that $\alpha\approx 0.1\, - \,0.2$ \cite{Smak99}, \cite{KL12}.
Only recently Hirose et al. \cite{Hirose14} showed that effects of convection at temperatures $\sim 10^4$\,K
increase $\alpha$ to values $\sim 0.1$. This might solve the problem of discrepancies between the MRI-calculated
and the observed value of $\alpha$ \cite{CKBL}. One has, however, to keep in mind that the simulations in question
have been performed in a so-called shearing box and their validity in a generic 3D case has yet to be demonstrated.

Another problem is related to cold discs such as quiescent dwarf nova discs \cite{Lasota01} or protostellar discs \cite{Balbus11}.
For the standard MRI to work, the degree of ionization in a weakly magnetized,
quasi-Keplerian disc must be sufficiently high to produce the instability that leads to a breakdown of laminar flow into turbulence
which is the source of viscosity driving accretion onto the central body. In cold discs the ionized fraction is very small and might be
insufficient for the MRI to operate. In any case in such a disc non-ideal MHD effects are always important. All these problems
still await their solution.

Finally, and very relevant to the subject of this Chapter there is the question of stability of discs in which the pressure
is due to radiation and opacity to electron scattering. According to theory such discs should be violently (thermally)
unstable but observations of systems presumed to be in this regime totally infirm this prediction. MRI simulations not only do
not solve this contradiction but rather reinforce it \cite{Jiang13}.

\subsection{The $\alpha$--prescription}

The $\alpha$--prescription \cite{SS73} is a rather simplistic description of the accretion disc
physics but before one is offered better and physically more reliable options its
simplicity makes it the best possible choice and has been the main source of progress
in describing accretion discs in various astrophysical contexts.

One keeps in mind that the accretion--driving viscosity is of magnetic origin, but one
uses an effective hydrodynamical description of the accretion flow.
The hydrodynamical stress tensor is (see e.g. \cite{LLFM87})
\begin{equation}
\label{eq:stress}
\tau_{r\varphi}= \rho \nu \frac{\partial v_{\varphi}}{\partial R}=\rho \frac{d\Omega}{d\ln R},
\end{equation}
where $\rho$ is the density, $\nu$ the kinematic viscosity coefficient and $v_\varphi$ the azimuthal velocity ($v_\varphi=R\Omega$).

In 1973 Shakura \& Sunyaev  proposed the (now famous) prescription
\begin{equation}
\label{eq:alphaP}
\tau_{r\varphi}=\alpha P,
\end{equation}
where $P$ is the total thermal pressure and $\alpha \leq 1$. This leads to
\begin{equation}
\label{eq:nualphaP}
\nu= \alpha c_s^2 \left[\frac{d\Omega}{d\ln R}\right]^{-1},
\end{equation}
where $c_s = \sqrt{P/\rho}$ is the isothermal sound speed and $\rho$ the density.
For the Keplerian angular velocity 
\begin{equation}
\label{eq: OmegaK}
\Omega=\Omega_K=\left(\frac{GM}{R^3}\right)^{1/2}
\end{equation}
this becomes
\begin{equation}
\label{eq:nualpha}
\nu=\frac{2}{3} \alpha c_s^2/\Omega_K.
\end{equation}
Using the approximate hydrostatic equilibrium Eq. (\ref{eq:mec_approx}) one can write this as
\begin{equation}
\label{eq:nu}
\nu \approx \frac{2}{3}\alpha c_s H.
\end{equation}

Multiplying the rhs of  Eq. (\ref{eq:stress}) by the ring length ($2\pi R$) and averaging over the (total) disc height one obtains the expression for the {\sl total torque}
\begin{equation}
\label{eq:stressH}
{\mathfrak T}= 2\pi R \Sigma \nu R\frac{d\Omega}{d\ln R},
\end{equation}
where
\begin{equation}
\label{eq:Sigma}
\Sigma=\int^{+\infty}_{-\infty}\rho\,dz .
\end{equation}
For a Keplerian disc 
\begin{equation}
\label{eq:torqueK}
{\mathfrak T}= 3\pi \Sigma \nu \ell_K,
\end{equation}
($\ell_K=R^2\Omega_K$ is the Keplerian \textsl{specific} angular momentum.)

The viscous heating is proportional to to $\tau_{r\varphi}(d\Omega/dR)$ \cite{LLFM87}.
In particular the viscous heating rate per unit volume is
\begin{equation}
\label{eq:visheat1}
q^+= -\tau_{r\varphi}\frac{d\Omega}{d\ln R},
\end{equation}
which for a Keplerian disc, using Eq. (\ref{eq:alphaP}), can be written as
\begin{equation}
\label{eq:qplusalphaP}
q^+=\frac{3}{2}\alpha \Omega_K P,
\end{equation}
and  the viscous heating rate per unit surface is therefore
\begin{equation}
\label{eq:vischeat2}
Q^+=\frac{{\mathfrak T\Omega^{\prime}}}{4\pi R}=\frac{9}{8} \Sigma \nu \Omega_K^2.
\end{equation}
(The denominator in the first rhs is $2\times 2\pi R$ taking into account the existence of two disc surfaces.)\\

\textsc{Additional reading:} References \cite{Balbus11}, \cite{BalbusHawley91} and \cite{Hirose14}.


\section{Geometrically thin Keplerian discs}
\label{sec:thin}

The 2D structure of geometrically thin, non--self-gravitating, axially symmetric accretion discs can be split into a 1+1 structure
corresponding to a hydrostatic vertical configuration and radial quasi-Keplerian viscous flow. The two 1D structures
are coupled through the viscosity mechanism transporting angular momentum and providing the local release
of gravitational energy.

\subsection{Disc vertical structure}
\label{subsec:vstruct.1}

The vertical structure can be treated as a one--dimensional star with two essential differences:
\begin{enumerate}
\item the energy sources are distributed over the whole height of the disc, while in a star there limited to the nucleus,
\item the gravitational acceleration \textsl{increases} with height because it is given by the tidal gravity of the accretor, while in stars it
decreases as the inverse square of the distance from the center.
\end{enumerate}

Taking these differences into account the standard stellar structure equations (see e.g. \cite{Prialnik}) adapted to the description of the disc vertical structure are listed
below.

\begin{itemize}
\item Hydrostatic equilibrium

The gravity force is counteracted by the force produced by the pressure gradient:
\begin{equation}
\label{eq:vertichydro}
\frac{dP}{dz}=\rho g_z,
\end{equation}
where $g_z$ is the vertical component (tidal) of the accreting body gravitational acceleration: 
\begin{equation}
\label{eq:vertacc}
g_z=\frac{\partial}{\partial z }\left[\frac{GM}{(R^2 + z^2)^{1/2}}\right]\approx \frac{GM}{R^2}\frac{z}{R}.
\end{equation}
The second equality follows from the assumption that $z\ll R$.
Denoting the typical (pressure or density) scale-height by $H$ the condition of geometrical thinness of the disc is $H/R\ll 1$
and writing $dP/dz \sim P/H$, Eq. (\ref{eq:vertichydro})
can be written as
\begin{equation}
\label{eq:mec_approx}
{H \over R}\approx {c_s\over v_K},
\end{equation}
where  $v_K=\sqrt{GM/R}$ is the Keplerian velocity and we made use of Eq. (\ref{eq:vertacc}). 
From Eq. (\ref{eq:mec_approx}) it follows that
\begin{equation}
\label{eq:dyntime}
\frac{H}{c_s} \approx \frac{1}{\Omega_K}= :t_{\rm dyn},
\end{equation}
where $t_{\rm dyn}$ is the dynamical time.\\

\item Mass conservation

In 1D hydrostatic equilibrium the mass conservation equation takes the simple form of
\begin{equation}
\label{eq:mass_cons}
\frac{d \varsigma}{dz} = 2 \rho.
\end{equation}
\medskip

\item Energy transfer - temperature gradient

\begin{equation}
\label{eq:nabla}
\frac{d\ln T}{dz} = \nabla \frac{d \ln P}{dz}.
\end{equation}
For radiative energy transport
\begin{equation}
\label{eq:nablarad}
\nabla_{\rm rad} = {\kappa_{\rm R} P F_{\rm z} \over 4 P_r c g_{\rm z}},
\end{equation}
where $P_r$ is the radiation pressure and $\kappa_{\rm R}$ the Rosseland mean opacity.
From Eqs. (\ref{eq:nabla}) and (\ref{eq:nablarad}) one recovers the familiar expression for
the radiative flux
\begin{equation}
\label{eq:radflux}
F_z= - \frac{16}{3}\frac{\sigma T^3}{\kappa_{\rm R}\rho}\frac{\partial T}{\partial z}=- \frac{4\sigma}{3\kappa_{\rm R}\rho}\frac{\partial T^4}{\partial z}
\end{equation}
($F_z$ is positive because the temperature decreases with $z$ so ${\partial T}/{\partial z}<0$).

The photosphere is at optical thickness $\tau\simeq 2/3$ (see Eq. \ref{t1}).
The boundary conditions are: $z = 0$, $F_z = 0$, $T = T_c$, $\varsigma = 0$ at
the disc midplane;  at the disc photosphere $\varsigma=\Sigma$ and
$T^4(\tau=2/3) = T^4_{\rm eff}$. For a detailed discussion of radiative transfer, temperature stratification and boundary conditions see Sect. \ref{subsect:rad}.

In the same spirit as Eq. (\ref{eq:mec_approx}) one can write Eq. (\ref{eq:radflux}) as 
\begin{equation}
\label{eq:rad_approx}
F_z\approx  \frac{4}{3}\frac{\sigma T_c^4}{\kappa_{\rm R}\rho H}= \frac{8}{3}\frac{\sigma T_c^4}{\kappa_{\rm R}\Sigma},
\end{equation}
where $T_c$ is the mid-plane (``central") disk temperature. Using the optical depth $\tau= \kappa_{\rm R}\rho H= (1/2) \kappa_{\rm R}\Sigma$, this can be written as
\begin{equation}
\label{eq:rad_approx2}
F_z(H) \approx  \frac{8}{3}\frac{\sigma T_c^4}{\tau}= Q^-,
\end{equation}
(see Eq. \ref{diff2} for a rigorous derivation of this formula).

\textsc{Remark 1.}
In some references (for example in \cite{FKR}) the numerical factor on the rhs is ``4/3" instead of ``8/3". This is due to a different definition of $\Sigma$: in our case it is $=2\rho H$, whereas in \cite{FKR} $\Sigma=\rho H$.
//
  
In the case of convective energy transport $\nabla=\nabla_{\rm conv}$. Because convection in discs is still not well understood (see, however, \cite{Hirose14}) there is no obvious choice for $\nabla_{\rm conv}$.
In practice a prescription designed by Paczy\'nski \cite{Paczynski69} for extended stellar envelope is used \cite{HMDLH} but this most probably does not represent very accurately what is happening in convective accretion discs \cite{CKBL}.

\item Energy conservation

Vertical energy conservation should have the form
\begin{equation}
\label{eq:strc}
\frac{dF_{\rm z}}{dz } = q^+ (z),
\end{equation}
where $q^+ (z)$ corresponds to viscous energy dissipation per unit volume.

\textsc{Remark 2.}
In contrast with accretion discs, stellar envelopes have ${dF_{\rm z}}/{dz }=0$

The $\alpha$ prescription does not allow deducing the viscous dissipation stratification ($z$ dependence), it just says that the vertically averaged viscous torque
is proportional to pressure. Most often one assumes therefore that 
\begin{equation}
\label{eq:strc2}
q^+ (z) = \frac{3}{2} \alpha \Omega_{\rm K} P (z),
\end{equation}
by analogy with Eq. (\ref{eq:qplusalphaP}) but such an assumption is chosen because of its simplicity and not because of some physical motivation. In fact MRI
numerical simulations suggest that dissipation is not stratified in the way as pressure \cite{Hirose14}.

\item The vertical structure equations have to be completed by the equation of state (EOS):
\begin{equation}
P = P_r + P_g= \frac{4\sigma }{3c}T^4 + \frac{{\cal R}}{\mu} \rho T ,
\label{eq:EOS}
\end{equation}
where ${\cal R}$ is the gas constant and $\mu$ the mean molecular weight, 
and an equation describing the mean opacity dependence on density and temperature.

\end{itemize}

\subsection{Disc radial structure}
\label{subsec:radial1}

\begin{itemize}
\item 
Continuity (mass conservation) equation has the form
\begin{equation}
\frac{\partial \Sigma}{\partial t} = - \frac{1}{R} \frac{\partial}{\partial
R} (R \Sigma v_{\rm r}) + \frac{S(R,t)}{2 \pi R}, 
\label{eq:consm}
\end{equation}
where $S(R,t)$ is the matter source (sink) term. 

In the case of an accretion disc in a binary system
\begin{equation}
\label{eq:source}
S(R,t)= \frac{\partial \dot{M}_{\rm ext}(R,t)}{
\partial R}
\end{equation}
represents the matter brought to the disc from the Roche lobe filling/mass losing (secondary) companion of the accreting object.
$\dot{M}_{\rm ext}\approx \dot{M}_{\rm tr}$, where $ \dot{M}_{\rm tr}$ is the mass transfer rate from the companion star.
Most often one assumes that the transfer stream delivers the matter exactly at the outer disc edge, but
although this assumption simplifies calculations it is contradicted by observations that suggest that the stream overflows the
disc surface(s).\\

\item Angular momentum conservation

\begin{equation}
\frac{\partial \Sigma \ell}{\partial t} = - \frac{1}{ R} \frac{\partial}{\partial
R} (R \Sigma \ell v_{\rm r}) + \frac{1}{ R} \frac{\partial}{\partial R}
\left(R^3 \Sigma \nu \frac{d\Omega}{dR} \right) + \frac{S_{\ell}(R,t)}{2\pi R}.
\label{eq:consj}
\end{equation}
This conservation equation reflects the fact that angular momentum is transported through the disc by a viscous stress $\tau_{r\varphi}=R \Sigma \nu {d\Omega}/{dR}$.
Therefore, if the disc is not considered infinite (recommended in application to real processes and systems) there must be somewhere a sink of this transported angular momentum $S_\ell (R,t)$. For binary semi-detached binary systems there is both a source (angular momentum brought in by the mass transfer stream form the stellar companion) and a sink (tidal interaction taking angular momentum back to the orbit). The two respective terms in the angular momentum equation can be written as
\begin{equation}
\label{eq:sourcej}
S_j(R,t)=\frac{ \ell_{\rm k}}{2 \pi R} \frac{\partial \dot{M}_{\rm ext}}{ \partial R} -
\frac{T_{\rm tid}(R)}{2 \pi R}.
\end{equation}

Assuming $\Omega=\Omega_K$, from Eqs. (\ref{eq:consm}) and (\ref{eq:consj}) one can obtain an diffusion equation for the surface density $\Sigma$:
\begin{equation}
\label{eq:eqdiff}
\frac{\partial \Sigma}{\partial t}=\frac{3}{R}\frac{\partial}{\partial R}\left\{ R^{1/2}  \frac{\partial}{\partial R}\left[ \nu \Sigma R^{1/2}\right]\right\}.
\end{equation}
Comparing with Eqs. (\ref{eq:consm}) one sees that the radial velocity induced by the viscous torque is
\begin{equation}
v_r= - \frac{3}{\Sigma R^{1/2}}\frac{\partial}{\partial R}\left[ \nu \Sigma R^{1/2}\right],
\label{eq:vr}
\end{equation}
which is an example of the general relation
\begin{equation}
\label{eq:vvisc}
v_{\rm visc}\sim \frac{\nu}{R}.
\end{equation}

Using Eq.(\ref{eq:nu}) one can write
\begin{equation}
\label{eq:vistime1}
t_{\rm vis} := \frac{R}{v_{\rm visc}}\approx \frac{R^2}{\nu}\approx\alpha^{-1}\frac{H}{c_s}\left(\frac{H}{R}\right)^{-2}.
\end{equation}
  
The relation between the viscous and the dynamical times is
\begin{equation}
t_{\rm vis}\approx  \alpha^{-1}\left(\frac{H}{R}\right)^{-2}\, t_{\rm dyn}.
\label{eq:vistime2}
\end{equation}
In thin ($H/R \ll 1$) accretion discs the viscous time is much longer that the dynamical time. In other
words, during viscous processes the vertical disc structure can be considered to be in hydrostatic equilibrium.

\item Energy conservation\\

The general form of energy conservation (thermal) equation can be written as:
\begin{equation}
\label{eq:energy}
\rho T\frac{d s}{dR}:=\rho T\frac{\partial s}{\partial t} +v_r\frac{\partial s}{\partial R}= q^+ - q^- + \widetilde q,
\end{equation}
where $s$ is the entropy density, $q^+$ and $q^-$ are respectively the viscous and radiative energy density, and
$\widetilde q$ is the density of external and/or radially transported energy densities.
Using the first law of thermodynamics
$Tds=dU +PdV$
one can write
\begin{equation}
\label{eq:firstlaw}
\rho T\frac{ds}{dt}=\rho\frac{d U}{d t} + P\frac{\partial v_r}{\partial r},
\end{equation}
where $U={{\Re} T_{\rm c} /\mu (\gamma - 1)}$.

Vertically averaging, but taking $T=T_c$, using Eq. (\ref{eq:consm}) and the thermodynamical relations from Appendix (for $\beta=1$) 
one obtains
\begin{equation}
\frac{\partial T_{\rm c}}{\partial t} +v_{\rm r} \frac{\partial T_{\rm c}}{\partial R}+
\frac{\Re T_{\rm c}}{\mu c_P} \frac{1}{R} \frac{\partial (R v_{\rm r})}{\partial R} 
= 2\frac{Q^+ -Q^-}{c_P \Sigma} + \frac{\widetilde Q}{c_P\Sigma} ,
\label{eq:heat}
\end{equation}
where $Q^+$ and $Q^-$ are respectively the heating and cooling rates per unit surface.
${\widetilde Q}=Q_{\rm out} + J$ with $Q_{\rm out}$ corresponding to energy contributions 
by the mass-transfer stream and tidal torques; $J(T,\Sigma)$ represent radial energy fluxes
that are a more or less ad hoc addition to the 1+1 scheme to which they do not belong since
it assumes that radial gradients ($\partial/\partial R$) of physical quantities can be neglected.

The viscous heating rate per unit surface can be written as (see Eq. \ref{eq:vischeat2})
\begin{equation}
Q^+=\frac{9}{8} \nu \Sigma \Omega_{\rm K}^2
\label{eq:qplus}
\end{equation}
while the cooling rate over unit surface (the radiative flux) is obviously
\begin{equation}
Q^- =\sigma T_{\rm eff}^4.
\label{eq:qminus}
\end{equation}
In thermal equilibrium one has
\begin{equation}
\label{eq:thermal_equi}
Q^+=Q^-.
\end{equation}

The cooling time can be easily estimated from Eq. (\ref{eq:thermal_equi}).
The energy density to be radiated away is $\sim \rho c_s^2$ (see Eqs \ref{eq:eqst} and \ref{eq:U}),
so the energy per unit surface is $\sim \Sigma c_s^2$ and the cooling (thermal) time is
\begin{equation}
\label{eq:thermal_time}
t_{\rm th}=\frac{\Sigma c_s^2}{Q^-}=\frac{\Sigma c_s^2}{Q^+}\sim \alpha^{-1}\Omega_K^{-1}= \alpha^{-1}t_{\rm dyn}.
\end{equation}
Since $\alpha < 1$, $t_{\rm th}>t_{\rm dyn}$  and during thermal processes the disc can be assumed to be in (vertical)
hydrostatic equilibrium.
  
For geometrical thin ($H/R\ll 1$) accretion discs one has the following hierarchy of the characteristic times
\begin{equation}
\label{eq:timeshare}
t_{\rm dyn} < t_{\rm th} \ll t_{\rm vis}.
\end{equation}
  
(This hierarchy is similar to that of characteristic times in stars: the dynamical is shorter than the thermal (Kelvin-Helmholtz)  and the
thermal is much shorter than the thermonuclear time.)
\end{itemize}

\subsection{Self-gravity}
\label{subsec:sg}

In this Chapter we are interested in discs that are not self-gravitating, i.e. in discs where the vertical hydrostatic equilibrium is maintained against the pull
of the accreting body's tidal gravity whereas the disc's self-gravity can be neglected. We will see now under what conditions this assumption
is satisfied.

The equation of vertical hydrostatic equilibrium can be written as 
\begin{equation}
\frac{1}{\rho}\frac{dP}{dz}= - g=\left(-g_z - g_s \right)= g_z\left(1+\frac{g_s}{g_z}\right)=:- g_z \left(1 + A\right),
\end{equation}
therefore self-gravity is negligible when $A \ll 1$.
Treating the disc as an infinite uniform plane (i.e. assuming the surface density does not vary too much with radius) one can write its self gravity as
$g_s=2\pi G\Sigma$, whereas the $z$-component of the gravity provided by the central body is $g_z=\Omega_K^2\,z$ (Eq. \ref{eq:vertacc}). Therefore evaluating $A$ at $z=H$
one gets
\begin{equation}
A_H:= \frac{g_s}{g_z}\bigg|_H = \frac{2\pi G \Sigma}{\Omega_K^2 H}.
\end{equation}
$A_H$ is related to the so-called Toomre parameter \cite{Toomre64} 
\begin{equation}
Q_T:=\frac{c_s\Omega}{\pi G\Sigma},
\end{equation}
widely used in the studies of gravitational stability of rotating systems, through $A_H \approx Q_T^{-1}$.
We will therefore express the condition of negligible self-gravity (gravitational stability) as
\begin{equation}
Q_T > 1.
\end{equation}
Using Eqs. (\ref{eq:mec_approx}), (\ref{eq:nu}) and (\ref{eq:amintK}) one can write the Toomre parameter as
\begin{equation}
Q_T=\frac{3c_s^3}{G\dot M},
\end{equation}
or as function of the mid-plane temperature $T=10^4\,T_4$
\begin{equation}
\label{eq:Qdisc}
Q_T\approx 4.6\times 10^{7}\frac{\alpha\, T^{3/2}_4}{m\,\dot m},
\end{equation}
where $m=M/\msun$. This shows that hot ionized ($T\gtrsim 10^4$) discs become self-gravitating for high accretor masses and
high accretion rates. Discs in close binary systems ($m \lesssim 30$) are never self-gravitating for realistic accretion
rates ($\dot m < 1000$, say) and even  in IMBH binaries (if they exist) (hot) discs would also be free of gravitational instability.
Around a supermassive black hole, however, discs can become self-gravitating quite close to the black hole. For example when
the black hole mass is $m=10^8$ a hot disc will become self-gravitating at $R/R_S\approx 100$, for $\dot m \sim 10^{-2}$.
In general, geometrically thin, non--self-gravitating accretion discs around supermassive black holes have very a limited radial extent.\\

\textsc{Additional reading:} References \cite{CollinZahn08}, \cite{Gammie01},\cite{Goodman03}, \cite{LinPringle87}, \cite{Paczynski78} and \cite{Toomre64}.

\subsection{Stationary discs}
\label{subsec:stationary}

In the case of stationary ($\partial/\partial t=0$) discs Eq. (\ref{eq:consm}) can be easily integrated giving
\begin{equation}
\label{eq:accrrate}
\dot M:=2\pi R \Sigma v_{\rm r},
\end{equation}
where the integration constant $\dot M$ (mass/time) is the \textsl{accretion rate}.

Also the angular momentum equation  (\ref{eq:consj}) can be integrated to give
\begin{equation}
\label{eq:amint1}
-2\pi R\Sigma v_r\ell  + 2\pi R^3\Sigma\nu\Omega'=const.
\end{equation}
Or, using Eq. (\ref{eq:accrrate}),
\begin{equation}
\label{eq:amint2}
-\dot M \ell   + {\mathfrak T}={const.},
\end{equation}
where the torque ${\mathfrak T}:= 2\pi R^3\Sigma\nu\Omega'$; (for a Keplerian disc ${\mathfrak T}= 3\pi R^2\Sigma\nu\Omega_K$).

Assuming that at the inner disc radius the torque vanishes one gets
$const. =- \dot M \ell_{\rm in}$, where $\ell_{\rm in}$ is the specific angular momentum at the
disc inner edge.
Therefore
\begin{equation}
\label{eq:amint3}
\dot M (\ell -\ell_{\rm in}) ={\mathfrak T}
\end{equation}
which is a simple expression of angular momentum conservation.

For Keplerian discs one obtains an important relation between viscosity and accretion rate
\begin{equation}
\label{eq:amintK}
\nu \Sigma =\frac{\dot M }{3\pi}\left[1 -\left(\frac{R_{\rm in}}{R}\right)^{1/2} \right].
\end{equation}
From Eqs. (\ref{eq:amintK}), (\ref{eq:qplus}),  (\ref{eq:qminus}), and the thermal equilibrium equation (\ref{eq:thermal_equi})
it follows that
\begin{equation}
\label{eq:teff}
\sigma T_{\rm eff}^4= \frac{8}{3}\frac{\sigma T_c^4}{\tau}=\frac{3}{8\pi}\frac{GM\dot M}{R^3}\left[1 -\left(\frac{R_{\rm in}}{R}\right)^{1/2} \right].
\end{equation}
\begin{figure}[t]
\includegraphics[scale=.3]{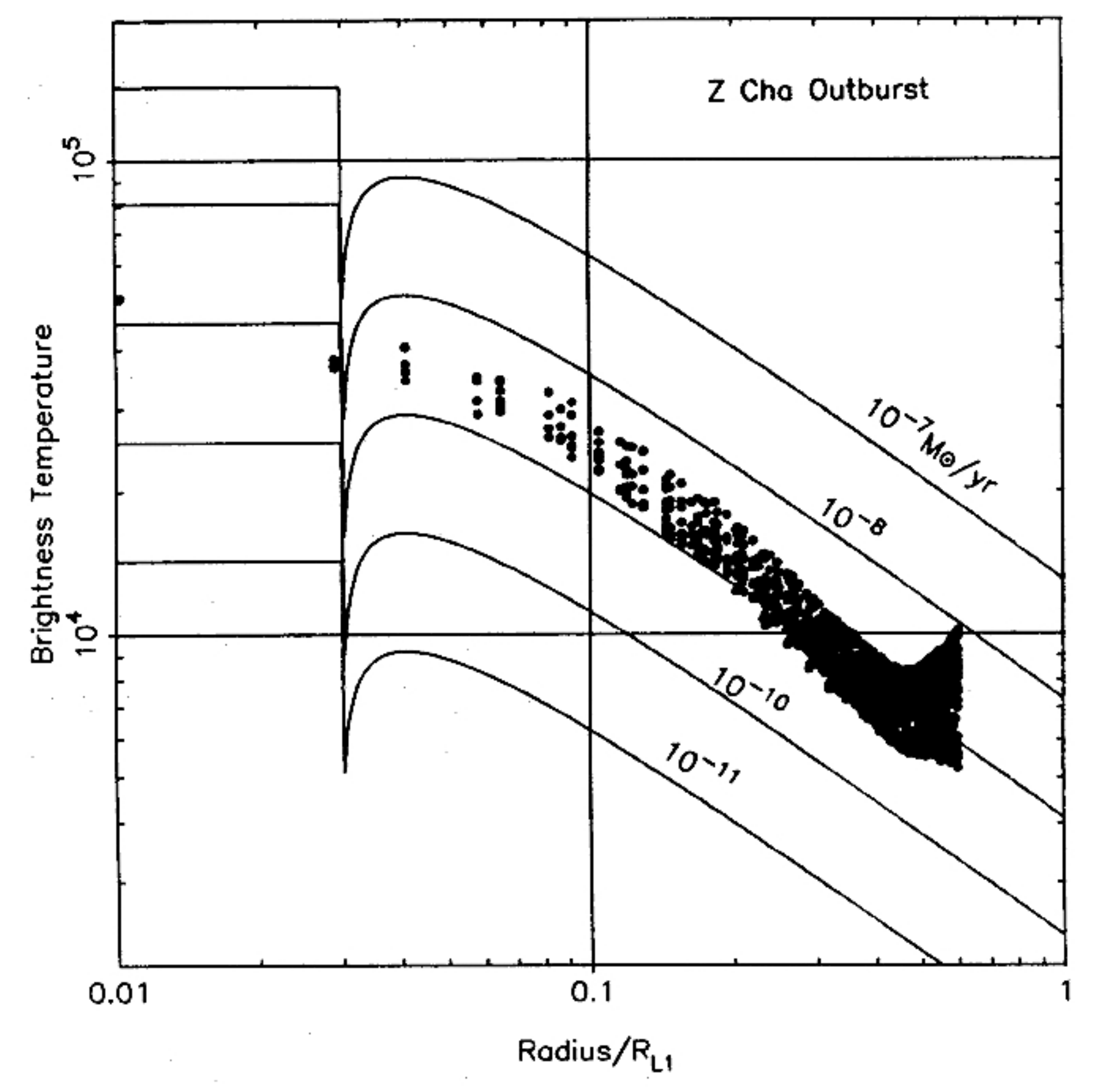}
\caption{The observed temperature profile of the accretion disc of the dwarf nova Z Cha in outburst. Near the outburst maximum such a disc is in quasi-equilibrium.
The observed profile, represented by dots (pixels), is compared with the theoretical profiles calculated from Eq. (\ref{eq:teff}) and represented by continuous lines.
Pixels with $R<0.03 R_{L1}$ correspond to to the surface of the accreting white dwarf whose temperature is 40 000 K. The accretion rate in the disc is $\approx 10^{-9}\msun \rm y^{-1}$.
[Figure 6 from \cite{HorneandCook85}].}
\label{fig:zchmax}      
\end{figure}
  
This relation assumes only that the disc is Keplerian and in thermal ($Q^+=Q^-$) and viscous ($\dot M=const.$) equilibrium. The viscosity coefficient is absent because of the thermal equilibrium assumption: in such a state the emitted radiation flux cannot contain information about the heating mechanism, it only says that such mechanism exists. Steady discs do not provide information about the viscosity operating in discs or the viscosity parameter $\alpha$. To get this information one must consider (and observe) time-dependent states of accretion discs.

From Eq. (\ref{eq:teff}) one obtains a universal temperature profile for \textsl{stationary Keplerian accretion discs}
\begin{equation}
\label{eq:temprofile}
T_{\rm eff}\sim R^{-3/4}.
\end{equation}
For an optically thick disc the observed temperature $T\sim T_{\rm eff}$ and $T\sim R^{-3/4}$ should be observed if
stationary, optically thick Keplerian discs exist in the Universe.  And vice versa, if they are observed, this proves that
such discs exist not only on paper. In 1985 Horne \& and Cook \cite{HorneandCook85} presented the observational proof
of existence of Keplerian discs when they observed the dwarf nova binary system ZCha during outburst (see Fig. \ref{fig:zchmax}).
\begin{figure}[t]
\includegraphics[scale=.3]{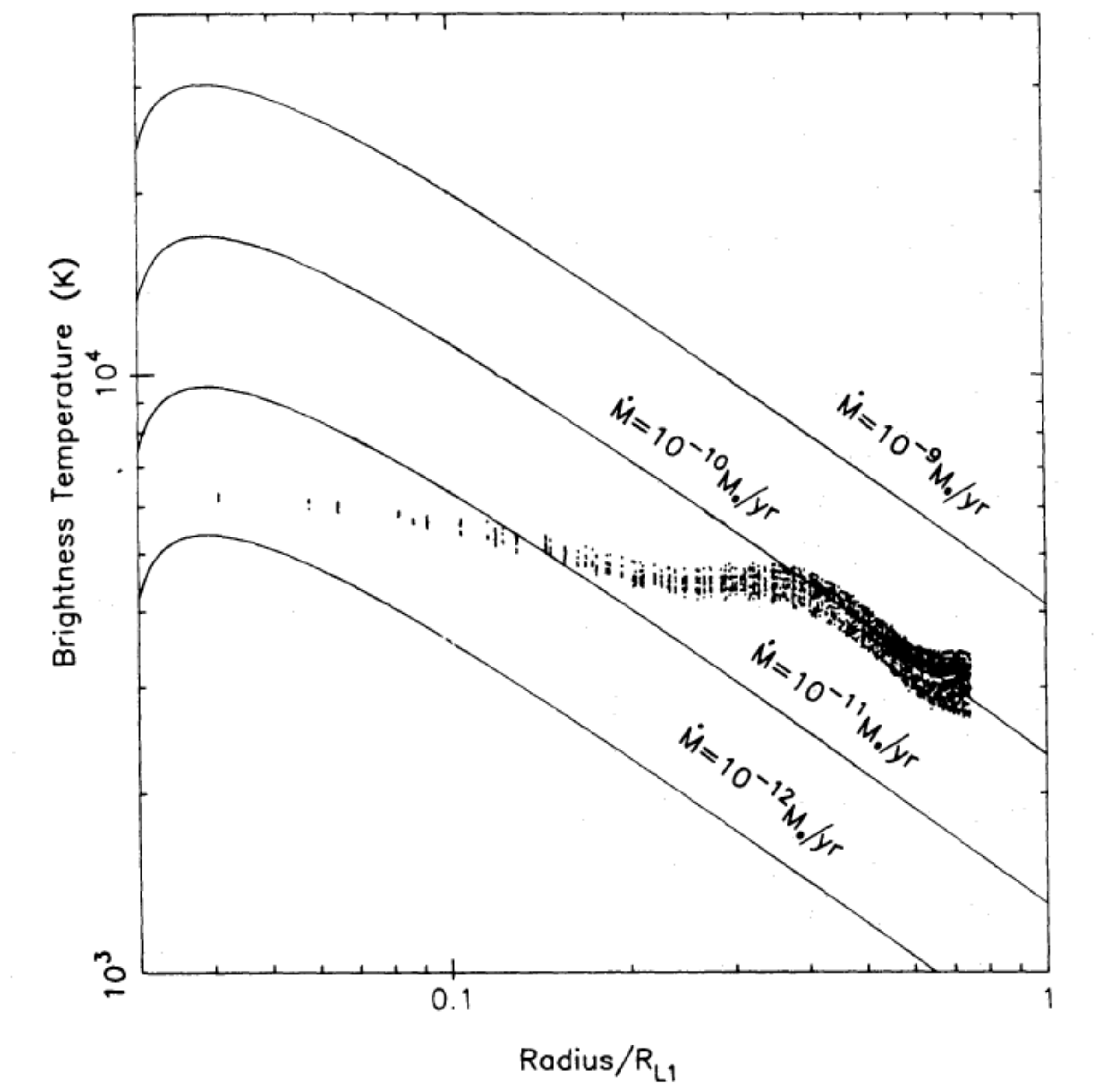}
\caption{The observed temperature profile of the accretion disc of the dwarf nova Z Cha in quiescence.
This one of the most misunderstood figures in astrophysics (see text). In quiescence the disc in \textsl{not} in equilibrium.
The flat temperature profile is \textsl{exactly}
what the disc instability model predicts: in quiescence the disc temperature \textsl{must} be everywhere lower than the critical temperature, 
but this temperature is almost independent of the radius (see Eq. \ref{eq:Tcrit}) .
[Figure 11 from \cite{Woodetal86}].}
\label{fig:zchaqn}       
\end{figure}

\subsubsection{Total luminosity}
\label{susect:totalL}

The total luminosity of a stationary, geometrically thin accretion disc, i.e. the sum of luminosities of its two surfaces, is
\begin{equation}
\label{eq:lum1}
2\int_{R_{\rm in}}^{R_{\rm out}}\sigma T_{\rm eff}^4\,2\pi RdR=  \frac{3GM\dot M}{2}\int_{R_{\rm in}}^{R_{\rm out}}\left[1 -\left(\frac{R_{\rm in}}{R}\right)^{1/2} \right]\frac{dR}{R^2}.
\end{equation}
For $R_{\rm out} \rightarrow \infty$ this becomes
\begin{equation}
\label{eq:lum2}
L_{\rm disc}=\frac{1}{2}\frac{GM\dot M}{R_{\rm in}}=\frac{1}{2}L_{\rm acc}.
\end{equation}
In the disc the radiating particles move on Keplerian orbits hence they retain half of the potential energy.
If the accreting body is a black hole this leftover energy will be lost (in this case, however, the non-relativistic formula of Eq.~\ref{eq:lum2}
does not apply -- see Eq.~\ref{eq:bindingE}.) In all the other cases the leftover energy will be released in the boundary layer,
if any, and at the surface of the accretor, from where it will be radiated away.
  
The factor ``3" in the rhs of Eq. (\ref{eq:teff}) shows that radiation by a given ring in the accretion disc does not come only from local
energy release. Indeed, in a ring between $R$ and $R+dR$ only
\begin{equation}
\label{eq:local}
\frac{GM\dot M dR}{2R^2}
\end{equation}
is being released, while
\begin{equation}
\label{eq:nonlocal}
2\times 2\pi R\, Q^+ dR=\frac{3GM\dot M}{2R^2}\left[1 -\left(\frac{R_{\rm in}}{R}\right)^{1/2} \right]{dR}
\end{equation}
is the total energy release. 
Therefore the rest
\begin{equation}
\label{eq:rest}
\frac{GM\dot M}{R^2}\left[1 -\frac{3}{2}\left(\frac{R_{\rm in}}{R}\right)^{1/2} \right]{dR}
\end{equation}
must diffuse out from smaller radii. This shows that viscous energy transport redistributes energy release in the disc.

\subsection{Radiative structure}
\label{subsect:rad}

Here we will show an example of the solution for the vertical thin disc structure
which exhibit properties impossible to identify when the structure is vertically
averaged. We will also consider here an irradiated disc -- such discs are present
in X-ray sources.

We write the energy conservation as :
\begin{equation}
{dF \over dz} = q^+(R,z),
\label{energy1}
\end{equation}
where $F$ is the vertical (in the $z$ direction) radiative flux and $q^+(R,z)$ is the viscous heating rate per unit volume.
Eq. (\ref{energy1})
states that an accretion disc is not in radiative equilibrium ($dF/dz\neq 0$), contrary to a
stellar atmosphere. For this equation to be solved, the function $q^+(R,z)$ must be known. As explained and discussed in Sect. \ref{subsec:vstruct.1} the viscous dissipation is often written as 
\begin{equation}
q^+(R,z)= \frac{3}{2} \alpha \Omega_{\rm K} P(z).
\label{voldiss}
\end{equation}
Viscous heating of this form has important implications for the structure of
optically thin layers of accretion discs and may lead to the creation of
coronae and winds. In reality it is an \textsl{an hoc} formula inspired by
Eq. (\ref{eq:qplusalphaP}). We don't know yet (see, however, \cite{CKBL}) how to describe the viscous
heating stratification in an accretion disc and Eq. (\ref{voldiss}) just
\textsl{assumes} that it is proportional to pressure. It is simple and convenient
but it is not necessarily true.

When integrated over $z$, the rhs of Eq. (\ref{energy1}) using Eq. (\ref{voldiss})
is equal to viscous dissipation per unit surface:
\begin{equation}
F^+={3 \over 2} \alpha \Omega_{\rm K} \int_0^{+\infty} P dz ,
\label{fvis}
\end{equation}
where $F^+=(1/2)Q^+$ because of the integration from $0$ to $+\infty$ while
$Q^+$ contains $\Sigma$ which is integrated from $-\infty$ to $+\infty$ (Eq. \ref{eq:Sigma}).

One can rewrite Eq. (\ref{energy1}) as
\begin{equation}
{dF\over d\tau} = - f(\tau ){F_{\rm vis} \over \tau_{\rm tot}},
\label{energy2}
\end{equation}
where we introduced a new variable, the optical depth $d\tau=-\kappa_{\rm R}
\rho dz$, $\kappa_{\rm R}$ being the Rosseland mean opacity and $\tau_{\rm tot}
= \int_0^{+\infty} \kappa_{\rm R} \rho dz$ is the total optical depth.
$f(\tau)$ is given by:
\begin{equation}
f(\tau) = {P \over \left(\int_0^{+\infty} P dz\right)} {\left(\int_0^{+\infty} \kappa_{\rm R}
\rho dz \right) \over \kappa_{\rm R} \rho}.
\label{deff}
\end{equation}
As $\rho$ decreases approximately exponentially, $f(\tau)$ is the ratio of
two rather well defined scale heights, the pressure and the opacity scale
heights, which are comparable, so that $f$ is of order of unity.

At the disc midplane, by symmetry, the flux must vanish: $F(\tau_{\rm tot})=0$,
whereas at the surface, ($\tau=0$)
\begin{equation}
F(0) \equiv \sigma T^4_{\rm eff}= F^+.
\label{fsurface}
\end{equation}
Equation (\ref{fsurface}) states that the total flux at the surface is equal
to the energy dissipated by viscosity (per unit time and unit surface). The
solution of Eq. (\ref{energy2}) is thus
\begin{equation}
F(\tau) = F^+ \left(1 - {\int_0^\tau f(\tau) d\tau \over \tau_{\rm tot}}\right),
\label{flux0}
\end{equation}
where $\int_0^{\tau_{\rm tot}} f(\tau) d\tau = \tau_{\rm tot}$. Given that
$f$ is of order of unity, putting $f(\tau) = 1$ is a reasonable
approximation. The
precise form of $f(\tau)$ is more complex, and is given by the functional
dependence of the opacities on density and temperature; it is of no
importance in this example. We thus take:
\begin{equation}
F(\tau) = F^+ \left(1 - {\tau\over \tau_{\rm tot}}\right).
\label{fluxd}
\end{equation}

To obtain the temperature stratification one has to solve the transfer
equation. Here we use the diffusion approximation
\begin{equation}
F(\tau) = {4 \over 3} {\sigma dT^4 \over d\tau} ,
\label{diff}
\end{equation}
appropriate for the optically thick discs we are dealing with. The
integration of Eq. (\ref{diff}) is straightforward and gives :
\begin{equation}
T^4(\tau) - T^4(0) = {3\over 4} \tau \left(1 - {\tau \over 2\tau_{\rm tot}}
            \right) T^4_{\rm eff}.
\label{t1}
\end{equation}

The upper (surface) boundary condition is:
\begin{equation}
T^4(0) = {1\over 2} T^4_{\rm eff} + T^4_{\rm irr},
\label{bcond2}
\end{equation}
where $T^4_{\rm irr}$ is the irradiation temperature, which depends on $r$,
the albedo, the height at which the energy is deposited and on the shape of
the disc. In Eq. (\ref{bcond2}) $T(0)$ corresponds to
the {\sl emergent} flux and, as mentioned above, $T_{\rm eff}$ corresponds to
the {\sl total} flux ($\sigma T^4_{\rm eff}=Q^+$) which explains the factor 1/2
in Eq (\ref{bcond2}). The temperature stratification
is thus :
\begin{equation}
T^4(\tau) = {3\over 4}T^4_{\rm eff}
             \left[\tau \left(1 - {\tau \over 2\tau_{\rm tot}}\right)
        + {2 \over 3}\right] + T^4_{\rm irr}.
\label{t2}
\end{equation}
For $\tau_{\rm tot} \gg 1$ the first
term on the rhs has the form familiar from the stellar atmosphere models in the
Eddington approximation.

In this case at $\tau=2/3$ one has $T(2/3) = T_{\rm eff}$.

Also for $\tau_{\rm tot} \gg 1$, the temperature
at the disc midplane is
\begin{equation}
T^4_{\rm c} \equiv T^4(\tau_{\rm tot}) =
         {3 \over 8} \tau_{\rm tot} T_{\rm eff}^4 + T^4_{\rm irr}.
\label{diff2}
\end{equation}
It is clear, therefore, that for the disc inner structure to be dominated by
irradiation and the disc to be isothermal one must have
\begin{equation}
{F_{\rm irr}\over \tau_{\rm tot}} \equiv {\sigma T^4_{\rm irr} \over
\tau_{\rm tot}} \gg F^+
\label{c1}
\end{equation}
and not just $F_{\rm irr} \gg F^+$ as is sometimes assumed. The
difference between the two criteria is important in LMXBs since, for
parameters of interest, $\tau_{\rm tot} \gtrsim 10^2 - 10^3$ in the outer disc
regions.

\subsection{Shakura-Sunyaev solution}
\label{subsec:SS}

In their seminal and famous paper Shakura \& Sunyaev \cite{SS73}, found power-law stationary solutions of the simplified version of the thin--disc equations presented in Sects. \ref{subsec:vstruct.1}, \ref{subsec:radial1} and \ref{subsec:stationary}. The 8 equations for the 8 unknowns $T_c$, $\rho$, $P$, $\Sigma$, $H$, $\nu$, $\tau$ and $c_s$ can be written as
\begin{equation}
\label{}
  \Sigma=2H\rho \ \ \ \  \ \ \  \ \ \ \ \ \ \ \ \ \ \ \ \ \ \ \ \ \ \ \ \ \ \ \ \ \ \ \ \ \ \ \ \ \ \ \ \ \ \ \ \ \ \ \ \ \ \ \ \ \ \ \ \ \ \ \ \ \rm (\i) \nonumber
\end{equation}
\begin{equation}
\label{}
 H=\frac{c_s R^{3/2}}{(GM)^{1/2}} \\ \ \ \ \ \ \ \ \ \ \ \ \ \ \ \ \ \ \ \ \ \ \ \ \ \ \ \ \ \ \ \ \ \ \ \ \ \ \ \ \ \ \ \ \ \ \ \ \ \ \ \ \ \ \ \ \ \ \ \rm (\i\i) \nonumber
\end{equation}
\begin{equation}
\label{}
c_s=\sqrt{\frac{P}{\rho}}\\ \ \ \ \ \ \ \ \ \ \ \ \ \ \ \ \ \ \ \ \ \ \ \ \ \ \ \ \ \ \ \ \ \ \ \ \ \ \ \ \ \ \ \ \ \ \ \ \ \ \ \ \ \ \ \ \ \ \ \ \ \ \ \ \ \ \rm (\i\i\i) \nonumber
\end{equation}
\begin{equation}
\label{}
P=\frac{{\cal R}\rho T}{\mu} + \frac{4\sigma}{3c}T^4 \\ \ \ \ \ \ \ \ \ \ \ \ \ \ \ \ \ \ \ \ \ \ \ \ \ \ \ \ \ \ \ \ \ \ \ \ \ \ \ \ \ \ \ \ \ \ \ \ \ \ \ \ \ \ \ \rm(\i v) \nonumber
\end{equation}
\begin{equation}
\label{}
\tau(\rho, \Sigma, T_c)=\kappa_R(\rho,T_c)\Sigma \\ \ \ \ \ \ \ \ \ \ \ \ \ \ \ \ \ \ \ \ \ \ \ \ \ \ \ \ \ \ \ \ \ \ \ \ \ \ \ \ \ \ \ \ \ \ \ \rm(v) \nonumber
\end{equation}
\begin{equation}
\label{}
\nu(\rho,\Sigma,T_c,\alpha)=\frac{2}{3}\alpha c_s H   \\ \ \ \ \ \ \ \ \ \ \ \ \ \ \ \ \ \ \ \ \ \ \ \ \ \ \ \ \ \ \ \ \ \ \ \ \ \ \ \ \ \ \ \ \ \ \ \ \ \rm(v\i) \nonumber
\end{equation}
\begin{equation}
\label{}
\nu \Sigma =\frac{\dot M }{3\pi}\left[1 -\left(\frac{R_0}{R}\right)^{1/2} \right] \\ \ \ \ \ \ \ \ \ \ \ \ \ \ \ \ \ \ \ \ \ \ \ \ \ \ \ \ \ \ \ \ \ \ \ \ \ \ \ \ \ \ \ \ \ \rm(v\i\i) \nonumber
\end{equation}
\begin{equation}
\label{}
\frac{8}{3}\frac{\sigma T_c^4}{\tau}=\frac{3}{8\pi}\frac{GM\dot M}{R^3}\left[1 -\left(\frac{R_0}{R}\right)^{1/2} \right] . \\ \ \ \ \ \ \ \ \ \ \ \ \ \ \ \ \ \ \ \ \ \ \ \ \ \ \ \ \ \ \ \rm(v\i\i\i) \nonumber
\end{equation}
Equations (\i) and (\i\i) correspond to vertical structure equations (\ref{eq:mass_cons}) and (\ref{eq:mec_approx}), Eq. (v\i\i) is the radial Eq. (\ref{eq:amintK}), while Eq. (v\i\i\i) connects vertical to radial equations. Eq. (\i\i\i) defines the sound speed, Eq. (\i v) is the equation of state and (v\i) contains the information about opacities. The viscosity $\alpha$ parametrization introduced in \cite{SS73} provides the closure of the 8 disc equations. Therefore they can be solved for a given set of $\alpha$, $M$, $R$ and $\dot M$.

Power-law solutions of these equations exist in physical regimes where the opacity can be represented in the  Kramers form $\kappa=\kappa_0\rho^{n}T^{m}$ and one of the two pressures, gas or radiation, dominates over the other. In \cite{SS73} three regimes have been considered: 
\begin{center}
$\left. a. \right)$ $P_r \gg P_g$ and $\kappa_{\rm es}\gg \kappa_{\rm ff}$\\
$\left. b. \right)$ $P_g \gg P_r$ and $\kappa_{\rm es}\gg \kappa_{\rm ff}$ \\
$\left. c. \right)$ $P_g \gg P_r$ and $\kappa_{\rm ff}\gg \kappa_{\rm es}$.\\
\end{center}
Regimes $\left. a. \right)$ and $\left. b. \right)$ in which opacity is dominated by electron scattering will be discussed in Sect. \ref{sect:advection}.
Here we will present the solutions of regime $\left. c. \right)$, i.e. we will assume that
\begin{equation}
\label{eq:kramers}
P_r=0 \ \ \ \ {\mathrm{and}} \ \ \ \ \kappa_R=\kappa_{\rm ff}=5\times 10^{24}\rho T_c^{-7/2}\,\rm cm^2g^{-1}.
\end{equation}
The solution for the surface density $\Sigma$, central temperature $T_c$ and the disc relative height (aspect ratio) are respectively
\begin{equation}
\label{eq:SigmaSS}
\Sigma=23\,\alpha^{-4/5}m^{1/4}R_{10}^{-3/4}{\dot M}_{17}^{7/10}f^{7/10}\, \rm g\,cm^{-2},
\end{equation}
\begin{equation}
\label{eq:TcSS}
T_c=5.8\times 10^4\,\alpha^{-1/5}m^{1/4}R_{10}^{-3/4}{\dot M}_{17}^{3/10}f^{3/10}\,\rm K,
\end{equation}
\begin{equation}
\label{eq:HoverRSS}
\frac{H}{R}= 2.4\,\times 10^{-2}\alpha^{-1/10}m^{-3/8}R_{10}^{1/8}{\dot M}_{17}^{3/20}f^{3/20},
\end{equation}
where $m=M/{\rm M_{\odot}}$, $R_{10}={R /(10^{10}\,\rm cm})$, $\dot M_{17}={\dot M/(10^{17}\,\rm g\,s^{-1}})$, and
$f~=~1~ -~ ({R_{\rm in}}/{R})^{1/2}$.
\begin{figure}[!ht]
\centering
\includegraphics[scale=0.6]{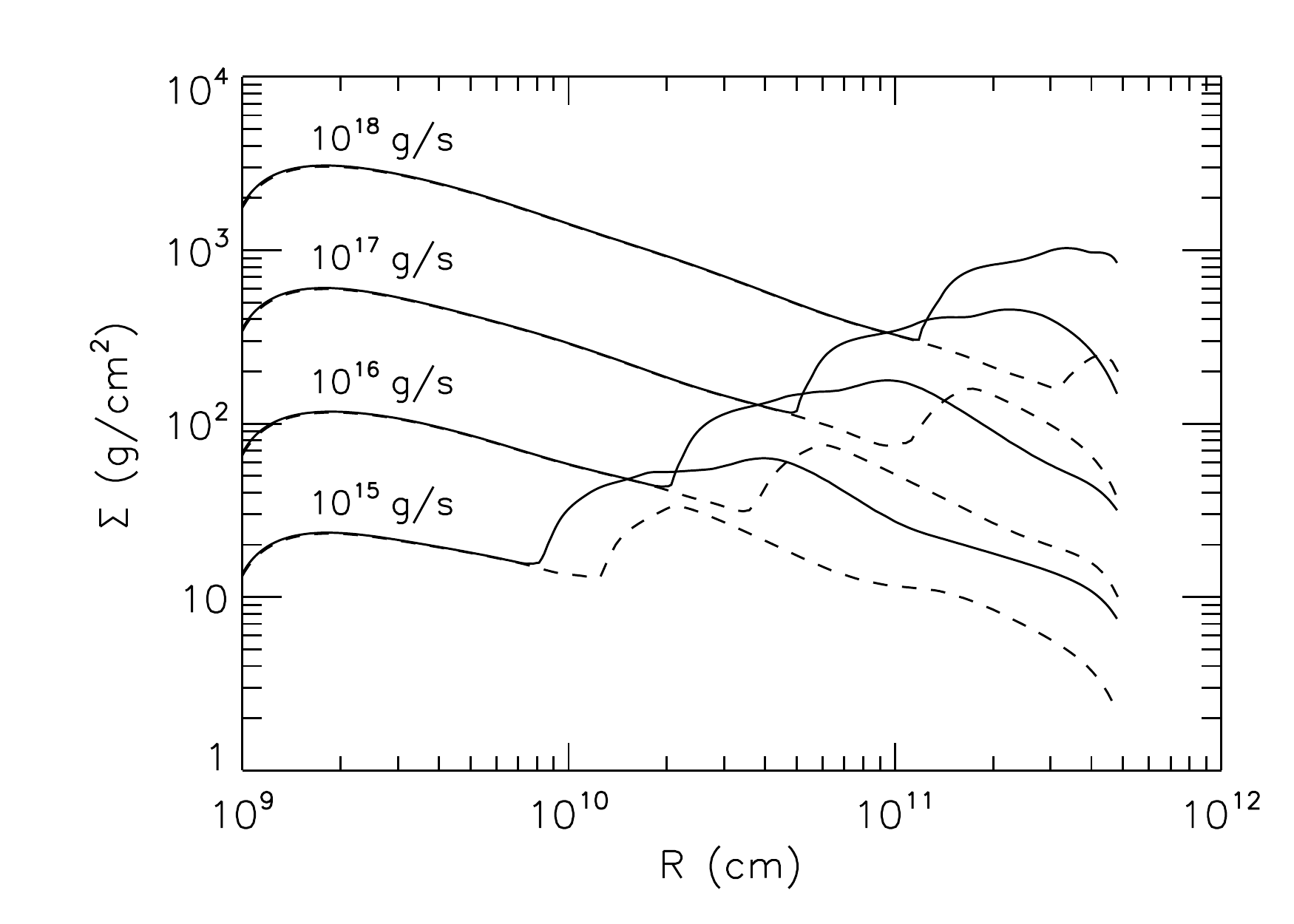}
  \caption{Stationary accretion disc surface density profiles for 4 values of accretion rate. From top to bottom: $\dot M =
10^{18}, 10^{17}, 10^{16}$ and $10^{15} \rm gs^{-1}$. $m = 10\msun$, $\alpha = 0.1$. The continuous line corresponds to the un-irradiated disc, the dotted lines to an irradiated configuration. The inner, decreasing segments of the continuous lines correspond to Eq. (\ref{eq:SigmaSS}). Dashed lines describe irradiated disc equilibria (see Sect. \ref{subsec:SXT}) [Figure 9 from \cite{DLHC}].} 
  \label{fig:tnsigprof}
\end{figure}

Although for a $10\,\msun$ black hole, say, Shakura-Sunyaev solutions (\ref{eq:SigmaSS}), (\ref{eq:TcSS}) and (\ref{eq:TcSS}) describe
discs rather far from its surface ($R\gtrsim 10^4\,R_S$) the regime of physical parameters it addresses, especially temperatures
around $10^4$K are of great importance for the disc physics because it is where accretion discs become thermally and viscously unstable.  
This instability triggers \textsl{dwarf nova} outbursts when the accreting compact object is a white dwarf and (soft) \textsl{X-ray transients}
in the case of accreting neutron stars and black holes. 

It is characteristic of the Shakura-Sunyaev solution in this regime that the three $\Sigma$, $T_c$ and $T_{\rm eff}$ radial profiles vary as $ R^{-3/4}$. 
(This implies that the optical depth $\tau$ is constant with radius -- see Eq. v\i\i\i.) For high accretion rates and small radii the
assumption of opacity dominated by free-free and bound-free absorption will break down and the solution will cease to be valid. We will come
to that later. Now we will consider the other disc end: large radii. 

One sees in Fig. \ref{fig:tnsigprof}, that for given stationary solution ($\dot M=const.$) the $R^{-3/4}$ slope of the $\Sigma$  profiles extends down only to a minimum value $\Sigma_{\rm min}(R)$
after which the surface density starts to increase. 
With the temperature dropping below $10^4$ K the disc plasma recombines and there is a drastic change in opacities leading to a thermal instability.

\textsc{Additional reading}: We have assumed that accretion discs are flat. This might not be true in general because accretion discs might be warped.
This has important and sometimes unexpected consequences; see e.g \cite{KingNixon13} \cite{Ogilvie99} and \cite{PP83}, and references therein.

\section{Disc instabilities}
\label{sec:DI}

In this section we will present and discuss the disc thermal and the (related) viscous instabilities.
First we will discuss in some detail the cause of the thermal instability due to recombination.

 \subsection{The thermal instability}
 \label{subsec:thermal_instability}

A disc is thermally stable if radiative cooling varies faster with
temperature than viscous heating. In other words
\begin{equation}
{d\ln \sigma T_{\rm eff}^4 \over d\ln T_{\rm c}} > {d\ln Q^+ \over
        d\ln T_{\rm c}}.
\label{stab}
\end{equation}
Using Eq. (\ref{diff2}) one obtains
\begin{equation}
{d\ln T_{\rm eff}^4 \over d\ln T_{\rm c}} =
4\left[1 - \left({T_{\rm irr} \over
    T_{\rm c}}\right)^{4}\right]^{-1} -
    {d\ln \kappa \over d\ln T_{\rm c}}.
\label{eq:cool}
\end{equation}
In a gas pressure dominated disc $Q^+ \sim \rho T\,H \sim \Sigma T \sim T_{\rm c}$ .
The thermal instability is due to a rapid change of opacities
with temperature when hydrogen begins to recombine. At high temperatures
${d\ln \kappa/d\ln T_{\rm c}}\approx - 4$ (see Eq. \ref{eq:kramers}). In the instability region, the
temperature exponent becomes large and positive ${d\ln \kappa/ d\ln T_{\rm
c}} \approx 7 - 10$, and in the end cooling is decreasing with temperature.
One can also see that irradiation by furnishing additional heat to the disc can stabilize an
otherwise unstable equilibrium solution (dashed lines in Fig. \ref{fig:tnsigprof}).

This thermal instability is at the origin of outbursts observed in discs around
black-holes, neutron stars and white dwarfs. Systems containing the first two
classes of objects are known as Soft X-ray transients (SXTs, where ``soft" relates
to their X-ray spectrum), while those containing white-dwarfs are called dwarf-novae
(despite the name that could suggest otherwise, nova and supernova outbursts
have nothing to do with accretion disc outbursts).

\subsection{Thermal equilibria: the \bS-curve}
\label{subsec:scurve}

We will first consider thermal equilibria of an accretion disc in which
heating is due only to local turbulence, leaving the discussion of the
effects of irradiation  to Section \ref{subsec:SXT}. We put therefore $T_{\rm
irr}=\widetilde Q=0$. Such an assumption corresponds to discs in cataclysmic variables
which are the best testbed for standard accretion disc models.
\begin{figure} [b]
\includegraphics[scale=0.30]{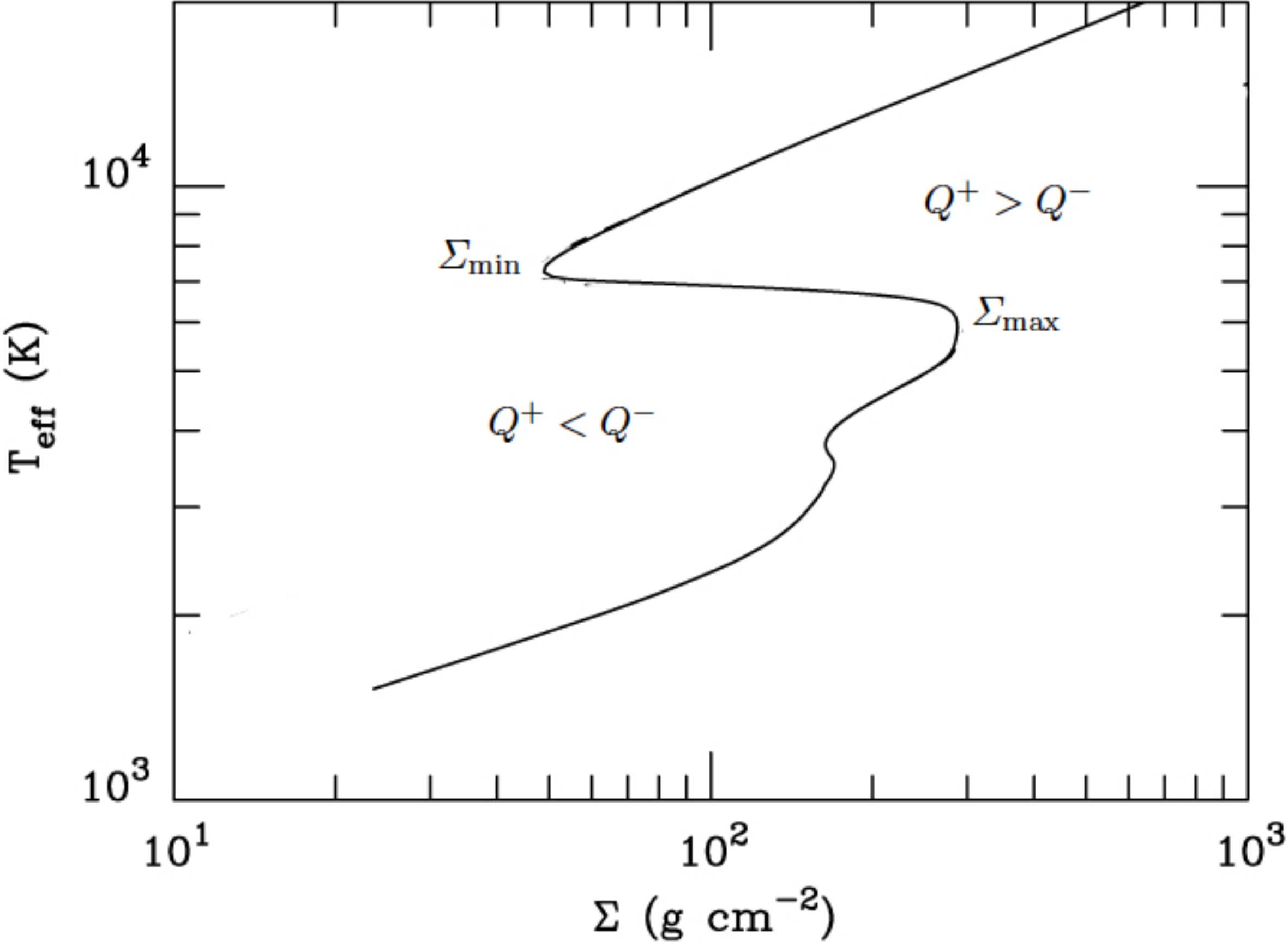}
  \caption{Thermal equilibria of a ring in an accretion discs around a  $m=1.2$ white dwarf. The distance from the center is
  $10^9$cm; accretion rate $6.66\times 10^{16}\rm g/s$. The solid line corresponds to $Q^+=Q^-$. $\Sigma_{\rm min}$ is
  the critical (minimum) surface density for a hot stable equilibrium; $\Sigma_{\rm max}$ the maximum surface density of
  a stable cold equilibrium.}
  \label{fig:scurve}
\end{figure}
The thermal equilibrium in the disc is defined by the equation $Q^-=Q^+$
(see Eq. \ref{eq:heat}), i.e. by
\begin{equation}
\sigma T_{\rm eff}^4=\frac{9}{8} \nu \Sigma \Omega_{\rm K}^2
\label{termeq}
\end{equation}
(Eq. \ref{eq:vischeat2}).
In general, $\nu$ is a function of density and temperature and in the following we will use
the standard $\alpha$--prescription Eq.  (\ref{eq:nualphaP}). The energy
transfer equation  provides a relation between the
effective and the disc midplane temperatures so that thermal equilibria
can be represented as a $T_{\rm eff}\left(\Sigma\right)$~ --~relation 
(or equivalently a $\dot M(\Sigma)$--relation). In the temperature range of interest ($10^3 \lesssim T_{\rm eff}\lesssim 10^5$)
this relation forms an {\bS} on the
($\Sigma, T_{\rm eff}$) plane as in Fig. \ref{fig:scurve}. The upper, hot branch corresponds to the Shakura-Sunyaev solution
presented in Section \ref{subsec:SS}. The two other branches correspond to solutions for cold discs -- along the middle branch 
convection plays a crucial role in the energy transfer.

Each point on the ($\Sigma, T_{\rm eff}$) \bS-curve represents an
accretion disc's thermal equilibrium at a given radius, i.e. a thermal equilibrium of a ring at radius $R$. In other words
each point of the \bS-curve is a solution of the $Q^+=Q^-$ equation. 
Points not on the \bS-curve correspond to solutions of Eq. (\ref{eq:heat})  \textsl{out
of thermal equilibrium}: on the left of the equilibrium curve cooling dominates over heating, $Q^+ < Q^-$; on the right 
heating over cooling $Q^+ > Q^-$. It is easy to see that a positive slope
of the $T_{\rm eff}(\Sigma)$ curve corresponds to \textsl{stable solutions}.
Indeed, a small increase of temperature of an equilibrium state (an upward perturbation) on the upper branch, say, 
will bring the ring to a state where $Q^+ < Q^-$ so it will cool down getting back to equilibrium. In a similar way
an downward perturbation will provoke increased heating bringing back the system to equilibrium. 

The opposite is happening along the \bS-curve's segment with negative slope as both temperature increase and
decrease lead to a runaway. The middle branch of the \bS-curve corresponds therefore to \textsl{thermally unstable}
equilibria.

A stable disc equilibrium can be represented only by a point on the lower, cold or
the upper, hot branch of the \bS-curve. This means that the surface
density in a stable cold state must be \textsl{lower} than the maximal value on the
cold branch: $\Sigma_{\rm max}$, whereas the surface density in the hot
stable state must be \textsl{larger} than the minimum value on this branch: $\Sigma_{\rm
min}$. Both these critical densities are functions of the viscosity
parameter $\alpha$, the mass of the accreting object, the distance from the
center and depend on the disc's chemical composition. In the case of solar composition the critical surface densities are
\begin{eqnarray}
\label{eq:Sigmacrit}
\Sigma_{\rm min}(R)&=&39.9~\alpha_{0.1}^{-0.80}~R_{10}^{1.11}~m^{-0.37}\, \rm g\,cm^{-2}\\
\Sigma_{\rm max}(R) &=&74.6~\alpha_{0.1}^{-0.83}~R_{10}^{ 1.18}~m_1^{-0.40}\, \rm g\,cm^{-2},
\end{eqnarray}
($\alpha=0.1\alpha_{0.1}$) and the corresponding effective temperatures ($T^+$ designates the temperature at $\Sigma_{\rm min}$,
$T^-$ at $\Sigma_{\rm max}$)
\begin{equation}
T_{\rm eff}^{+} =6890~R_{10}^{-0.09}~M_1^{0.03}\,\rm K 
\label{eq:Tcritplus}
\end{equation}
\begin{equation}
T_{\rm eff}^{-} =5210~R_{10}^{-0.10}~M_1^{ 0.04}\,\rm K.
\label{eq:Tcrit}
\end{equation}
The critical effective temperatures are practically independent of the mass and radius because they characterize
the microscopic state of disc's matter (e.g. its ionization).
On the other hand the critical accretion rates depend very strongly on radius:
\begin{eqnarray}
\dot{M}_{\rm crit}^{+}(R)&=&8.07\times10^{15}~\alpha_{0.1}^{-0.01}~R_{10}^{2.64}~M_1^{-0.89}\,\rm g\,s^{-1} \label{eq:Mcritplus}\\
\dot{M}_{\rm crit}^{-}(R)&=&2.64\times10^{15}~\alpha_{0.1}^{0.01}~R_{10}^{ 2.58}~M_1^{-0.85}\,\rm g\,s^{-1}.
\label{eq:Mcrit}
\end{eqnarray}

A stationary accretion disc in which there is a ring with effective temperature contained between the critical values of
Eq. (\ref{eq:Tcrit}) and (\ref{eq:Tcritplus}) cannot be stable. Since the effective temperature and the surface density both decrease with radius, the
stability of a disc depend on the accretion rate and the disc size (see Fig. \ref{fig:tnsigprof}). For a given
accretion rate a stable disc cannot have an outer radius larger than the value corresponding to Eq. (\ref{eq:Sigmacrit}).
  
A disc is stable if the rate (mass-transfer rate in a binary system) at which mass is brought to its outer edge ($R\sim R_d$)  is larger than the critical accretion rate
at this radius $\dot{M}_{\rm crit}^{+}(R_d)$.
  
In general, the accretion rate and the disc size are determined by mechanisms and conditions that are exterior to the accretion process
itself. In binary systems, for instance, the size of the disc is determined by the masses of the system's components and its orbital period
while the accretion rate in the disc is fixed by the rate at which the stellar companion of the accreting object loses mass, which in turn
depends on the binary parameters and the evolutionary state of this stellar mass donor. Therefore the knowledge of the orbital period 
and the mass-transfer rate should suffice to determine if the accretion disc in a given interacting binary system is stable. Such knowledge
allows testing the validity of the model as we will show in the next section.

\subsubsection{Dwarf nova and X-ray transient outbursts}
\label{subsub:dnoutb}
\begin{figure} [t]
\includegraphics[scale=0.27]{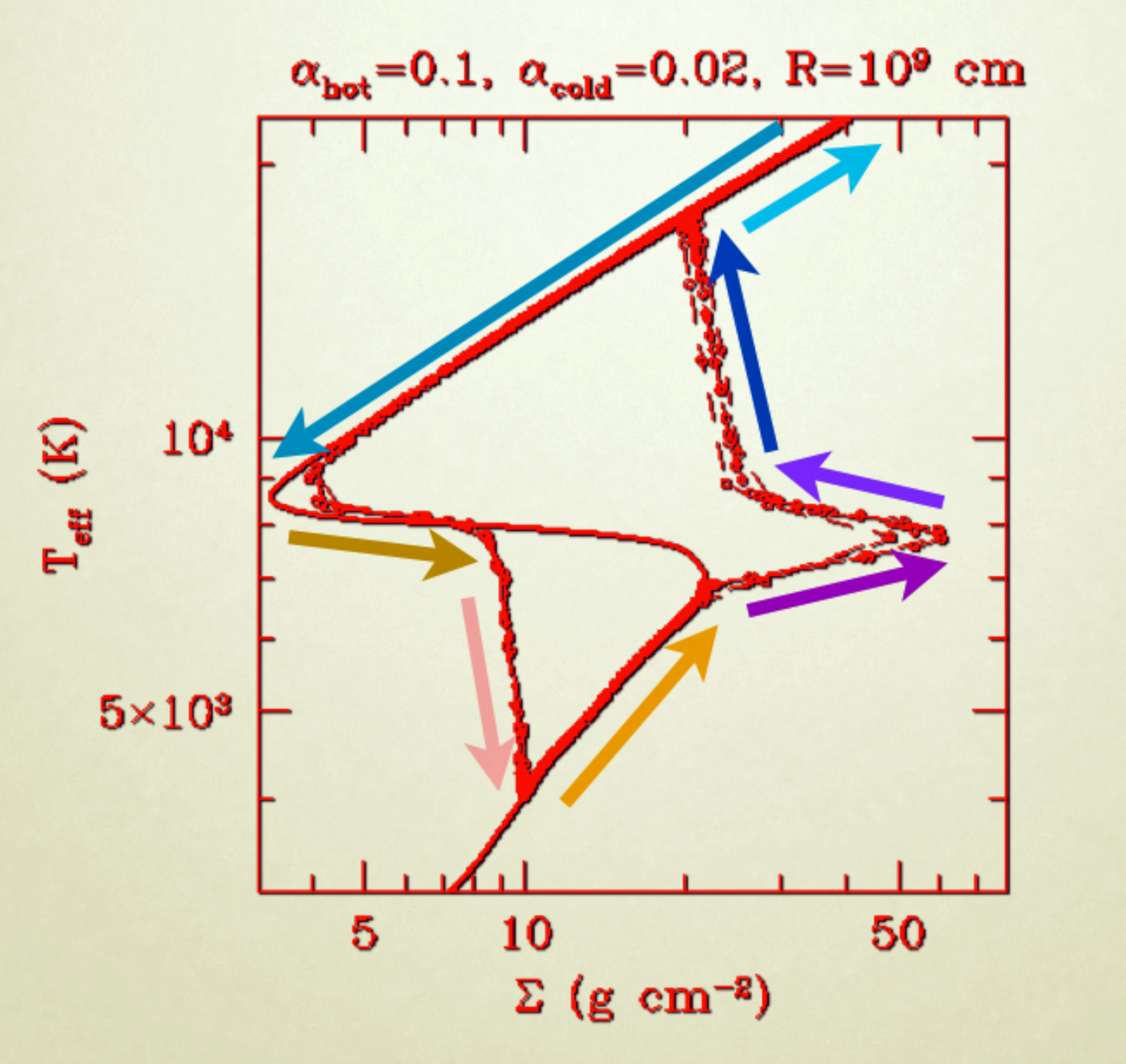}
  \caption{Local limit cycle of the state of disc ring at $10^9$ cm during a dwarf nova outbursts.
  The arrows show the direction of motion of the system in the $T_{\rm eff}(\Sigma)$ plane.
  The figure represents results of the disc instability model numerical simulations. As required
  by the comparison of the model with observations the values of the viscosity parameter $\alpha$
  on the hot and cold branches are different.
   [Figure adapted from \cite{MHS99}]}
  \label{fig:cycle}
\end{figure}
\begin{itemize}
\item Local view: the limit cycle
\end{itemize}
Let us first describe what is happening during outbursts with a disc's ring. Its states are represented by a point moving
in  the $\Sigma - T_{\rm eff}$ plane as shown on Fig. \ref{fig:scurve} which represents accretion disc states at $R=10^9$ cm
(the accreting body has a mass of $1.2\msun$). 
To follow the states of a ring during the outburst let us start with an unstable equilibrium
state on the middle, unstable branch and let us perturb it by increasing its temperature, i.e. let us shift it upwards in the $T_{\rm eff}(\Sigma)$ plane. 
As we have already learned,  points out of the \bS-curve correspond to solutions
out of thermal equilibrium and in the region to the right of the \bS-curve
heating dominates over cooling. The resulting runaway temperature increase is represented by the point moving up and reaching (in a thermal time) a quasi--equilibrium state on the hot and stable branch. It is only a \textsl{quasi}--equilibrium because the equilibrium state has been assumed to lie on the middle branch which corresponds to a lower temperature (and lower accretion rate -- see Eq. \ref{eq:teff}). Trying to get to its proper equilibrium the ring will cool down and move towards lower temperatures and surface densities along the upper equilibrium branch (in a viscous time). But the hot branch ends at $\Sigma_{\rm min}$, i.e. at a temperature higher (and surface density lower) than required so the ring will never reach its equilibrium state. Which is not surprising since this state is unstable. Once more the ring will find itself out of thermal equilibrium but this time in the region where cooling dominates over heating. Rapid (thermal-time) cooling will bring it to the lower cool branch. There, the temperature is lower than required so the point representing the ring will move up towards $\Sigma_{\rm max}$ where it will have to interrupt its (viscous-time) journey having reached the end of equilibrium states before getting to the right temperature. It will find itself out of equilibrium where heating dominated over cooling so it will move back to the upper branch.

Locally, the state of a ring performing a \textsl{limit cycle } on the $\Sigma$--$T_{\rm eff}$ plane, moves in viscous time on the stable \bS-curve branches and in a thermal time between them when the ring is out of thermal equilibrium. The states on the hot branch correspond to outburst maximum and the subsequent decay whereas the quiescence correspond to moving on the cold branch. Since the viscosity is much larger on the hot than on the cold branch, the quiescent is much longer than the outburst phase. The full outburst behaviour can be understood only by following the whole disc evolution (Figs. \ref{fig:rise} \& \ref{fig:decay}).

\subsection{Irradiation and black--hole X-ray transients}
\label{subsec:SXT}
We will present the global view of thermal-viscous disc outbursts for the case of X-ray transients. The main difference between accretion discs
in dwarf novae and these systems is the X-ray irradiation of the outer disc in the latter. Assuming that the irradiating X-rays are emitted
by a point source at the center of the system, one can write the irradiating flux as 
\begin{equation}
\label{eq:C}
\sigma T^4_{\rm irr}={\mathcal C}\frac{L_X}{4\pi R^2} \hskip0.5truecm \mathrm{with} \hskip0.5truecm L_X= \eta\, \mathrm{min} \left(\dot M_{\rm in}, \dot M_{\rm Edd}\right)c^2,
\end{equation}
where ${\mathcal C}=10^{-3}{\mathcal C_3}$, $\eta$ is the radiative efficiency (which can be $\ll 0.1$ for ADAFs - see below), $\dot M_{\rm in}$ the accretion rate at the inner disc's edge. Since the physics and geometry of X-ray self-irradiation in accreting black-black hole systems is still
unknown, the best we can do is to parametrize our ignorance by q constant $\mathcal C$ that observations suggest is $\sim 10^{-3}$. Of course one
should keep in mind that in reality $\mathcal C$ might not be a constant \cite{Esin}.

\begin{figure} [!]
\includegraphics[width=\textwidth]{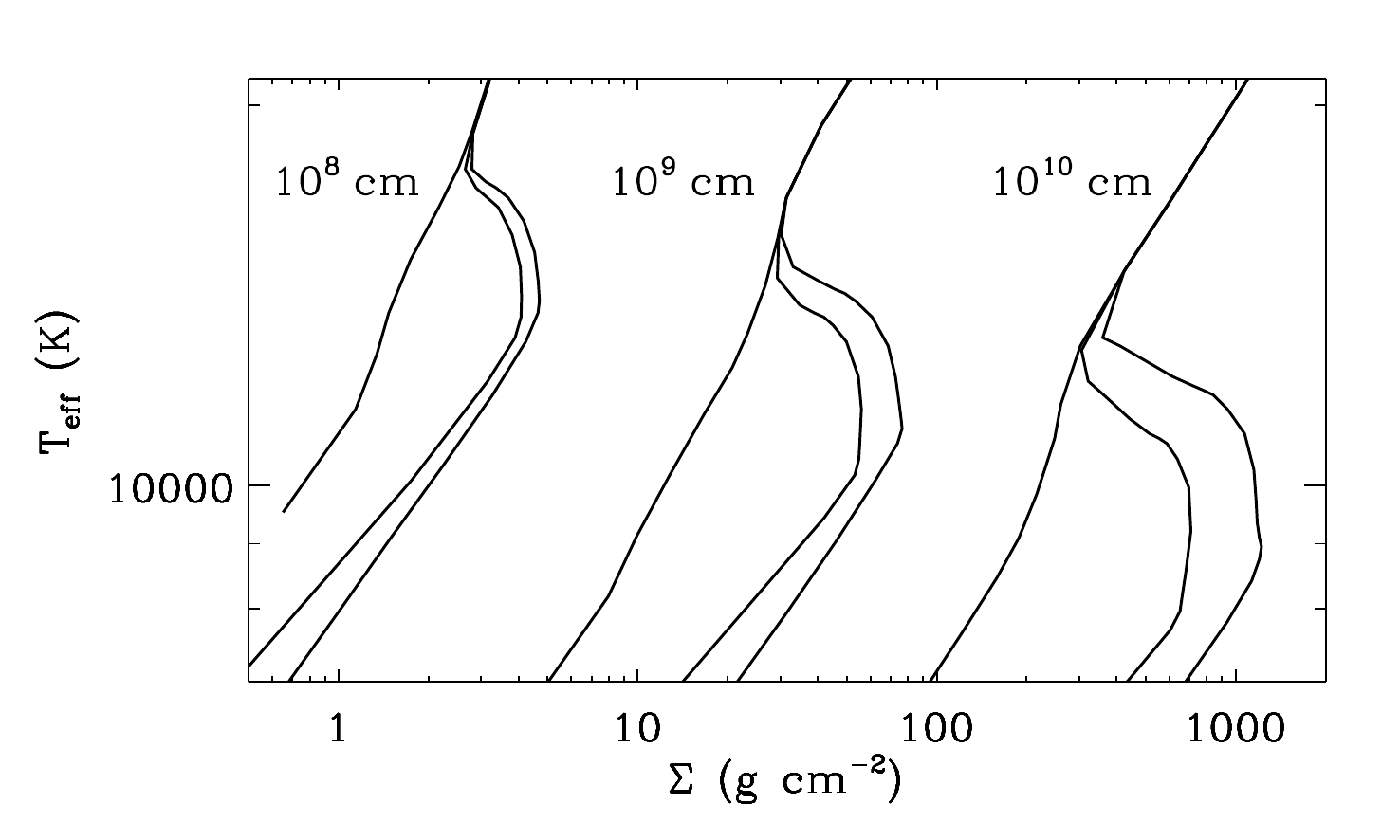}
  \caption{Example S-curves for a pure helium disk with varying
irradiation temperature $T{\rm irr}$.  The various
sets of S-curves correspond to radii $R=10^6$, $10^9$ and $10^{10}$\,cm. For each radius, the irradiation temperature $T_{\rm
  irr}$ is 0\,K, 10 000\,K and 20 000~K. $\alpha=0.16$. The instable branch disappears for high irradiation temperatures.
  [From \cite{LDK}. Reproduced with permission from Astronomy \& Astrophysics, \copyright ESO]}
  \label{fig:Sirr}
\end{figure}

Because the viscous heating is $\sim \dot M/R^3$ there always exists a radius $R_{\rm irr}$ for which $\sigma T^4_{\rm irr}> Q^+=\sigma T^4_{\rm eff}$.
If $R_{\rm irr} < R_d$, where $R_d$ is the outer disc radius, the outer disc emission will be dominated by reprocessed X-ray irradiation and the structure
modified as shown in Sect. \ref{subsect:rad}. Irradiation will also stabilize outer disc regions (Eq. \ref{eq:cool} and Fig. \ref{fig:Sirr}) allowing larger discs for a given
accretion rate (see Fig. \ref{fig:tnsigprof}). 

Irradiation modifies the critical values of the hot disc parameters:
\begin{eqnarray}
\Sigma_{\rm irr}^{+}  &=&72.4
                    ~{\cal C}_{-3}^{-0.28}
                    ~\alpha_{0.1}^{-0.78}
                    ~R_{11}^{0.92}
                    ~M_1^{-0.19}\, \rm g\,cm^{-2} \\
T_{\rm eff}^{\rm irr,+} &=&2860
                    ~{\cal C}_{-3}^{-0.09}
                    ~\alpha_{0.1}^{0.01}
                    ~R_{11}^{-0.15}
                    ~M_1^{ 0.09}\,\rm K \\
\dot{M}_{\rm irr}^{+} &=&2.3 \times 10^{17}
                    ~{\cal C}_{-3}^{-0.36}
                    ~\alpha_{0.1}^{ 0.04}
                    ~R_{11}^{2.39}
                     ~M_1^{-0.64}\,\rm g\,s^{-1}.
\label{eq_h-irrad} 
\end{eqnarray}

As we will see in a moment, irradiation also strongly influences the shape of outburst's light-curve.

\subsection*{* Rise to outburst maximum}

\begin{figure} [!]
\centering
\includegraphics[scale=0.40]{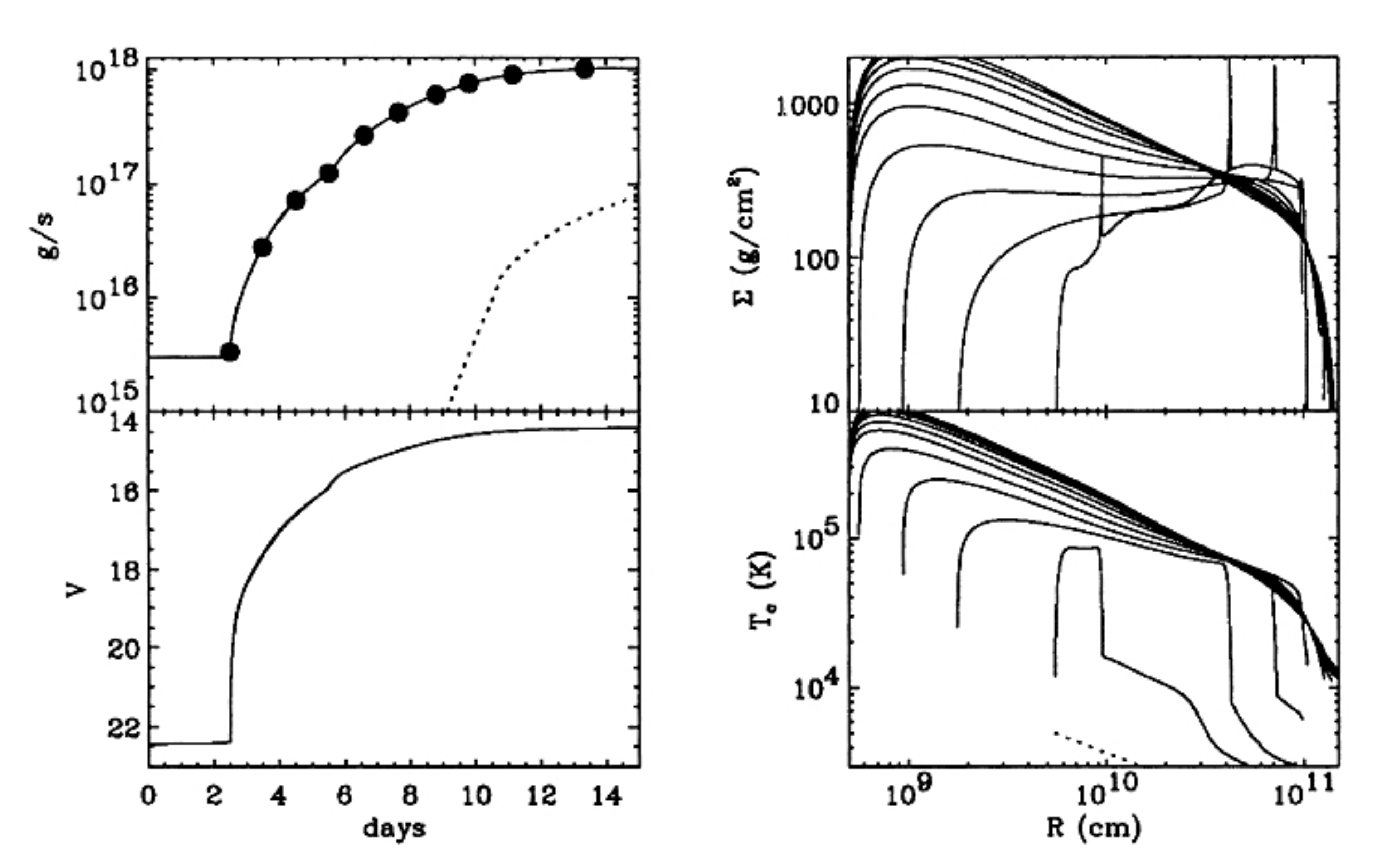}
  \caption{The rise to outburst described in Sect.
\ref{subsec:SXT}. The upper left panel shows $\dot{M}_{\rm in}$ and
$\dot{M}_{\rm irr}$ (dotted line); the bottom left panel shows the $V$
magnitude. Each dot corresponds to one of the $\Sigma$ and $T_{\rm c}$
profiles in the right panels. The heating front propagates outwards. The
disc expands during the outburst due to the  angular momentum transport
of the material being accreted. At $t\approx 5.5$~days the thin disc
reaches the minimum inner disc radius of the model. The profiles close
to the peak are those of a steady-state disc ($\Sigma\propto T_{\rm
c}\propto R^{-3/4}$). [From \cite{DHL01}. Reproduced with permission 
from Astronomy \& Astrophysics, \copyright ESO]}
 \label{fig:rise}
\end{figure}

During quiescence the disc's surface density, temperature and accretion rate are everywhere (at all radii) on the cold  branch,
below their respective critical values $\Sigma_{\rm max}(R)$, $T_{\rm eff}^{-}$ and $\dot{M}_{\rm crit}^{-}(R)$.
It is important to realize that in quiescence the disc is \textsl{not} steady: $\dot M \neq const.$
Matter transferred from the stellar companion accumulates in the disc and is redistributed by viscosity. 
The surface density and temperature increase (locally, this means that the solution moves up along the lower branch of the \bS--curve) 
finally reaching their critical values. In Fig. \ref{fig:rise} this happens at $\sim 10^{10}$\,cm. The disc parameters entering the unstable regime triggers an outburst.
In the local picture this corresponds to leaving the lower branch of the
\bS-curve. The next `moment' (in a thermal time is
represented in the left panels of Figure \ref{fig:rise}. This is when a
large contrast forms in the midplane temperature profile and when a
surface-density spike is already above the critical line. The disc is
undergoing a thermal runaway at $r\approx 8\times 10^9$ cm. The midplane
temperature rises to $\sim 70000$ K. This raises the viscosity which
leads to an increase of the surface-density and a heating fronts start 
propagating inwards and outwards in the disc as seen in Fig. \ref{fig:rise}. 
In this model the disc is truncated at an inner radius $R_{\rm in}\approx 6\times 10^9$cm
so the inwards propagating front quickly reaches the inner disc radius 
with no observable effects. It is the outwards propagating heating front that
produces the outburst by heating up the disc and redistributing the mass
and increasing the surface density behind it because  it is also a compression front.

One should stress here that two ad hoc elements must be added to the model
for it to reproduce observed outbursts of dwarf novae and X-ray transients.
\begin{itemize}
\item \textsc{Viscosity.}
 First, if the increase in viscosity were
due only to the rise in the temperature through the speed of sound ($\nu
\propto c_s^2$, see Eq. \ref{eq:nualpha}) the resulting outbursts would have
nothing to do with the observed ones. To reproduce observed outbursts
one increases the value of $\alpha$ when a given ring of the disc gets to the hot branch.
Ratios of hot--to--cold $\alpha$ of the order of 4 are used to describe
dwarf nova outburst. Although in the outburst model the $\alpha$ increase is an ad hoc assumption, recent MRI
simulations with physical parameters corresponding to dwarf nova discs show
an $\alpha$ increase induced by the appearance of convection \cite{Hirose14}.
\item \textsc{Inner truncation}.
Second, as mentioned already, the inner disc is assumed to be truncated
in quiescence and during the rise to outburst. Although such truncation
is implied and/or required by observations, its physical origin is still uncertain.
The inner part of the accretion flow is of course not empty but supposed
to form a $\dot M=const.$ ADAF (see Sect. \ref{sect:advection}).
\end{itemize}

In our case (Fig. \ref{fig:rise}), the heating front reaches the outer
disc radius. This corresponds to the largest outbursts. 
Smaller-amplitude outbursts are produced when the front does not
reach the outer disc regions. In an inside-out outburst\footnote{X--ray transient outbursts 
are always of inside-out type. In dwarf novae both inside-out and
outside-in outbursts are observed and result from calculations\cite{Lasota01}.} the
surface-density spike has to propagate uphill, against the surface-density
gradient because just before the outburst $\Sigma\sim R^{1.18}$ -- roughly
parallel to the critical surface-density. Most of the mass is
therefore contained in the outer disc regions. A heating front will be able
to propagate if the post-front surface-density is larger than
$\Sigma_{\rm min}$ -- in other words, if it can bring successive rings
of matter to the upper branch of the \bS-curve. If not, a {\sl cooling
front} will appear just behind the $\Sigma$ spike, the heating front
will die-out and the cooling front will start to propagate inwards (the
heating-front will be `reflected').

The difficulty inside-out fronts encounter when propagating is due to
angular-momentum conservation. In order to move outwards the
$\Sigma$-spike has to take with it some angular momentum because the
disc's angular momentum increases with radius. For this reason inside-out
front propagation induces a strong \textsl{outflow}. In order for matter to be
accreted, a lot of it must be sent outwards. That is why during an
inside-out dwarf-nova outburst only $\sim 10\%$ of the disc's mass is
accreted onto the white dwarf. In X-ray transients irradiation facilitates
heating front propagation (and disc emptying during decay -- see next section).

The arrival of the heating front at the outer disc rim does not end
the  rise to maximum.  After the whole disc is
brought to the hot state,  a surface density (and accretion rate)
`excess' forms in the outer disc.  The accretion rate in the inner
disc corresponds to the critical one but  is much higher near the
outer edge. While irradiation keeps  the disc hot the excess diffuses
inwards until the accretion rate is  roughly constant. During this
last phase of the rise to outburst maximum  $\dot M_{\rm in}$
increases by a factor of 3: 
\begin{equation}
\dot{M}_{\rm max}\approx 3 \dot{M}_{\rm irr}^{+}\approx 7.0 \times 10^{17}{\mathcal  C}_{-3}^{-0.36}~R_{d,11}^{2.39}~m^{-0.64}\,\rm g\,s^{-1}.
\label{eq:Mdotmax}
\end{equation}

Irradiation has little influence on
the actual  vertical structure in this region
and  $T_{\rm c}\propto\Sigma\propto   R^{-3/4}$, as in a
non-irradiated steady disc. Only in the outermost   disc regions does
the vertical structure becomes   irradiation-dominated, i.e. isothermal.

\subsection*{* Decay}

\begin{figure} [t]
\centering
\includegraphics[scale=0.40]{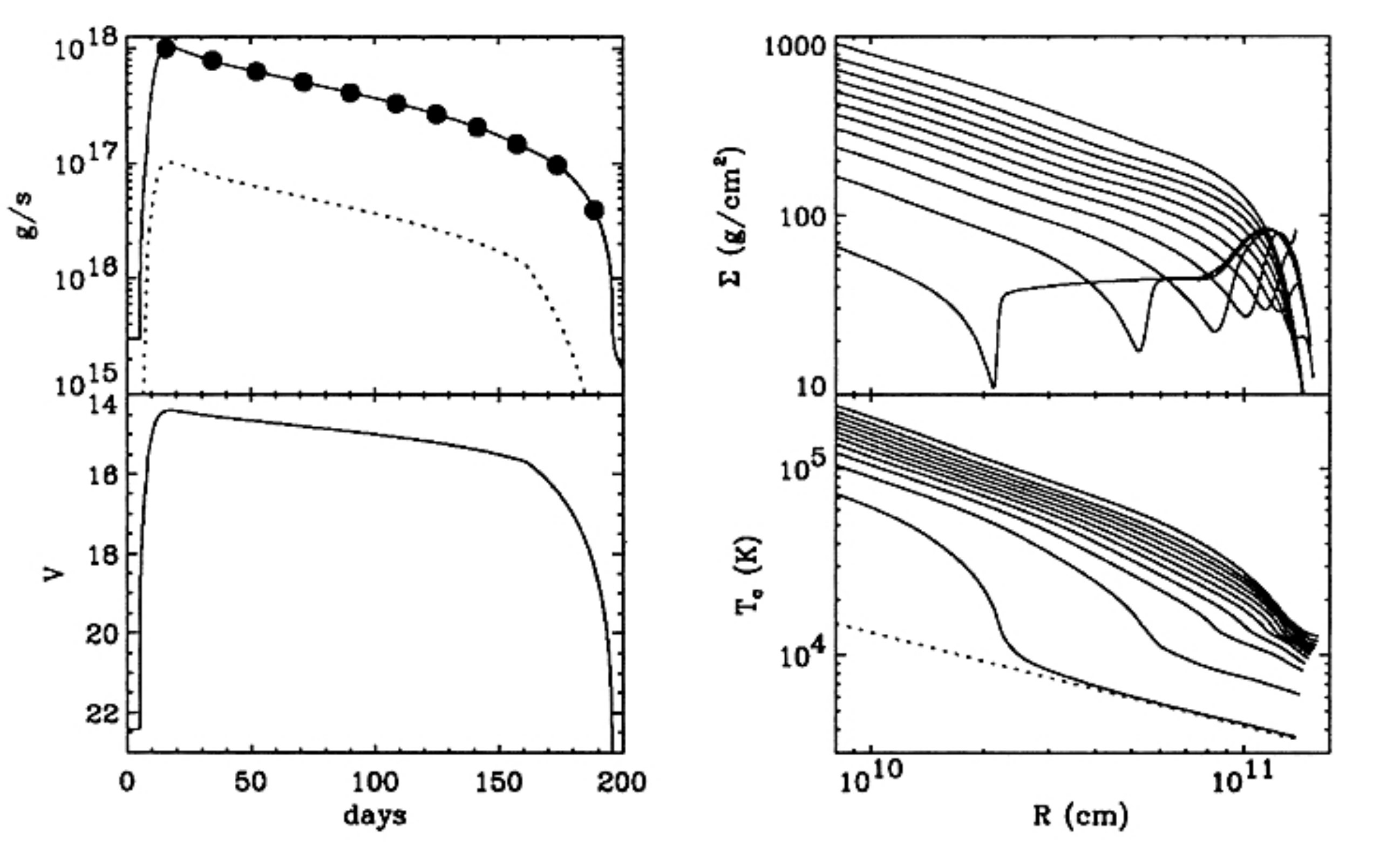}
  \caption{Decay from outburst peak. The decay is controlled by irradiation until
evaporation sets in at $t\approx170$~days ($\dot{M}_{\rm
in}=\dot{M}_{\rm evap}(R_{\rm min})$). This cuts off irradiation and the
disc cools quickly. The irradiation cutoff happens before the cooling
front can propagate through most of the disc, hence the
irradiation-controlled linear decay ($t\approx 80-170$~days) is not very
visible in the lightcurve. $T_{\rm irr}$ (dotted line) is shown for the
last temperature profile. [From \cite{DHL01}. Reproduced with permission from Astronomy \& Astrophysics, \copyright ESO]}
  \label{fig:decay}
\end{figure}

Fig. \ref{fig:decay} shows the sequel to what was described in Fig.
\ref{fig:rise}. In general the decay from the outburst peak of an irradiated disc
can be divided into three parts: 

\begin{itemize}

\item First, X-ray irradiation of the outer disc inhibits cooling-front
propagation. But since the peak accretion rate is much higher than the
mass-transfer rate,\footnote{The peak luminosity is $\sim 3\dot{M}_{\rm irr}^{+}(R_d)$;
and the for the disc to be unstable the mass-transfer rate must be lower than the critical rate: $\dot M_{\rm tr}< \dot{M}_{\rm irr}^{+}(R_d)$.} 
the disc is drained by viscous accretion of matter.

\item Second, the accretion rate becomes too low for the X-ray
irradiation to prevent the cooling front from propagating. The
propagation speed of this front, however, is controlled by irradiation.

\item Third, irradiation plays no role and the cooling front
switches off the outburst on a local thermal time-scale.
\end{itemize}

\subsubsection*{`Exponential decay'}

In Fig. \ref{fig:decay} the ``exponential decay" the phase lasts until roughly day 80-100. At the
outburst peak the accretion rate is almost exactly constant with radius; the disc is quasi-stationary.
The subsequent evolution is self-similar: the disc's radial structure
evolves through a sequence of quasi-stationary ($\dot M(r)=const$)
states. Therefore $\nu \Sigma \sim \dot{M}_{\rm in}(t)/3\pi$ and the total
mass of the disk is thus
\begin{equation}
M_{\rm d}=\int 2\pi R\Sigma dR \propto \dot{M}_{\rm in} \int {2\over 3} {r
\over \nu} dr .
\label{md}
\end{equation}

At the outburst peak the whole disc is wholly ionized and
except for the outermost regions its structure is very well represented
by a Shakura-Sunyaev solution. In such discs, as well as in irradiation
dominated discs, the viscosity coefficient satisfies the relation $\nu \propto
T \propto \dot{M}^{\beta/\left(1+\beta\right)}$.
In hot Shakura-Sunyaev discs $\beta = 3/7$ (Eq. \ref{eq:TcSS}), and  in irradiation
dominated discs $\beta$ = 1/3 (Eq. \ref{eq:C}). During the first decay phase the outer
disc radius is almost constant so that using Eq. (\ref{md})
the disc-mass evolution can be written as:
\begin{equation}
\frac{{\rm d} M_{\rm d}}{{\rm d} t}=-\dot{M}_{\rm in} \propto M_{\rm d}^{1+\beta}
\label{eq:expo}
\end{equation}
showing that $\dot{M}_{\rm in}$ evolves almost exponentially, as long
as $\dot{M}_{\rm in}^\beta$ can be considered as constant (i.e. over
about a decade in $\dot{M}_{\rm in}$.
`Exponential' decays in the DIM are only approximately exponential.

The quasi--exponential decay is due to {\sl two} effects:
\begin{enumerate}
\item X-ray irradiation keeps the
disc ionized, preventing cooling-front  propagation,
\item tidal torques
keep the outer disc radius roughly constant.
\end{enumerate}

\subsubsection*{`Linear' decay}

The second phase of the decay begins when a disc ring cannot remain in
thermal equilibrium. Locally this corresponds to a fall onto the cool
branch of the \bS-curve. In an irradiated disc this happens when the
central object does not produce enough X-ray flux to keep the $T_{\rm
irr}(R_{\rm out})$ above $\sim 10^4$~K. A cooling front appears and
propagates down the disc at a speed of $v_{\rm front}\approx \alpha_{\rm
h} c_s$.

In an irradiated disc, however, the transition between the hot and cold
regions is set by $T_{\rm irr}$ because a cold branch exists only for
$T_{\rm irr}\lesssim 10^4$~K. In an irradiated disc a cooling front can
propagate inwards only down to the radius at which $T_{\rm irr}\approx
10^4$~K, i.e. as far as there is a cold branch to fall onto.  Thus the
decay is still irradiation-controlled. The hot region remains close to
steady-state but its size shrinks $R_{\rm hot}\sim \dot{M}_{\rm
in}^{1/2}$ (as can be seen in Eq.~\ref{eq:C} with $T_{\rm irr}(R_{\rm
hot})= \mathrm{const}$).

\subsubsection*{Thermal decay}

In the model shown in Fig. \ref{fig:decay} irradiation is
unimportant after $t\gtrsim 170-190$~days because $\eta$ becomes
very small for $\dot{M}_{\rm in}<10^{16}$~g$\cdot$s$^{-1}$ when an ADAF forms.
The cooling front thereafter propagates freely inwards, on a thermal time
scale. In this particular case the decrease of irradiation is caused by
the onset of evaporation at the inner edge which lowers the efficiency. In general there
is always a moment at which $T_{\rm irr}$ becomes less than $10^4$~K;
evaporation just shortens the `linear' decay phase.

\subsection{Maximum accretion rate and decay timescale}

Now we will see that there are two observable properties of X-ray transients that, when related one to to the other, provide informations and constraints on the physical properties of the outbursting system.
The first is the maximum accretion rate $\dot M_{\rm max}$ (Eq. \ref{eq:Mdotmax}).
The second is the decay time of the X--ray flux: as we have seen, disk irradiation by the central X--rays traps the disk in the hot, high state, and only allows a decay of $\dot M$ on the hot--state viscous timescale.
This is
\begin{equation}
t \simeq \frac{R^2}{3\nu}
\label{eq:t}
\end{equation}
which using Eq. (\ref{eq:nualpha}) gives
\begin{equation}
t \simeq \frac{(GMR)^{1/2}}{3\alpha c_s^2}.
\label{eq:tvisc1}
\end{equation}
Taking the critical midplane temperature $T_{\rm c}^{+}\approx16000\,\rm K$ one gets
for the decay timescale
\begin{equation}
t\approx 32   \ m^{1/2} R_{d,11}^{1/2} \alpha_{0.2}^{-1} \,\rm days,
\label{eq:tvisc}
\end{equation} 
where $\alpha_{0.2}=\alpha/0.2$. 
Eliminating $R$ between (\ref{eq:Mdotmax}) and (\ref{eq:tvisc}) gives the
accretion rate through the disk at the start of the outburst as
\begin{equation}
\dot{M}=5.4 \times 10^{17}\ m^{-3.03} \left(t_{30}\alpha_{0.2}\right)^{4.78} \rm g\,s^{-1},
\label{eq:mdot}
\end{equation} 
with $t = 30\, t_{30}\,{\rm d}$. Assuming an efficiency of $\eta$ of 10\%, the corresponding  luminosity  is
\begin{equation}
L=5.0 \times 10^{37}\ \eta_{0.1} m^{-3.03} \left(t_{30}\alpha_{0.2}\right)^{4.78} \rm erg\,s^{-1}
\label{eq:luminosity1}.
\end{equation}

\subsection{Comparison with observations}

\subsubsection{Sub-Eddington outbursts}
\label{ssubsec:subEdout}

The peak luminosities of most of the soft X-ray transients are sub-Eddington.
Eq. (\ref{eq:mdot}) can be written using the Eddington ratio $m:=\dot M/\dot M_{\rm Edd}$ as
\begin{equation}
\dot m = 0.42 \eta_{0.1}(\alpha_{0.2}t_{30})^{4.78}m^{-4.03}.
\label{eddratio}
\end{equation}
This  equation shows that the outburst peak will be sub--Eddington only if the outburst decay time is relatively short or the accretor (black hole) mass is high, i.e. the observed decay timescale is
\begin{equation}
 t \lesssim 50\, \eta_{0.1}^{-0.21}\alpha_{0.2}^{-1}m^{0.84}~{\rm d},
\label{eq:subedd}
\end{equation}
in good agreement with the compilation of X--ray transients outburst durations found in \cite{YanYu14}. This shows that  the standard value of  efficiency $\eta_{0.1}\simeq1$,  and the value $\alpha_{0.2}\simeq1$ deduced from observations of dwarf novae, give the correct order of magnitude for the decay timescale of X--ray transients (from $\approx 3$ days to $\approx 300$ days). This equation also implies that black hole transients should have longer decay timescales than neutron star transients, all else being equal. Yan and Yu \cite{YanYu14} find that outbursts last on average $\approx 2.5\times$ longer in  black hole transients than in neutron star transients thus confirming this conclusion.

For sub--Eddington outbursts Eq.~(\ref{eq:luminosity1})  gives a useful  relationship between distance $D$, bolometric flux $F$ and outburst decay time $t$, 
\begin{equation}
D_{\rm Mpc} \simeq 1.0\, m^{-1.5}\left(\frac{\eta_{0.1}}{ F_{12}}\right)^{1/2}(\alpha_{0.2}t_{50})^{2.4},
\label{eq:distance}
\end{equation}
where $D = D_{\rm Mpc}\,{\rm Mpc}$ and  $F = 10^{-12}F_{12}\,{\rm erg\, s^{-1}\,cm^{-2}}$; $F=L/4\pi D^2$ and $t = 50\, t_{50}\,{\rm d}$.

Eq. (\ref{eq:distance}) shows that distant ($D > 1 \mathrm{Mpc}$) X-ray sources exhibiting variability typical of soft X-ray transients cannot contain black holes with masses superior
to stellar masses \cite{LKD}.

\subsubsection{Observational tests}
\begin{figure} [!]
\centering
\includegraphics[width=\textwidth]{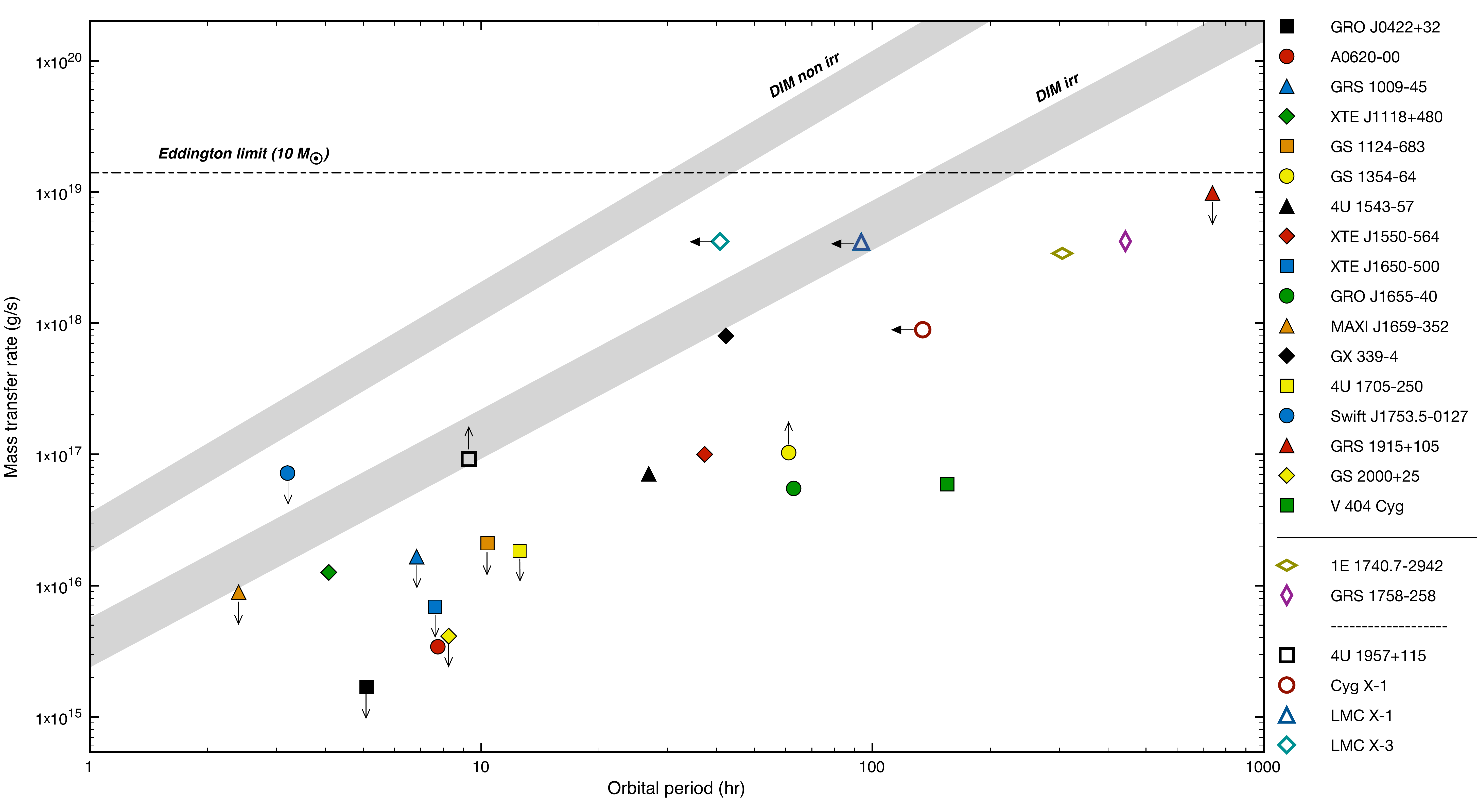}
  \caption{Mass transfer rate as a function of the orbital period for SXTs with black holes. The transient and persistent sources have been marked with respectively filled and open 
	symbols. The shaded grey areas indicated `DIM irr' and `DIM non irr' represent the separation between persistent (above) and transient systems (below) according to the disc instability model when, respectively, irradiation is taken into account and when it is neglected. The horizontal dashed line indicates the Eddington accretion rate for a $10 \msun$ black hole. All the upper limits on the mass transfer rate are due to lower limits on the recurrence time. The upper limits on the mass transfer rate of 4U 1957+115 and GS 1354-64 result from lower limits on the distance to the sources. The three left closed arrows do not indicate actual upper limits on the orbital period of Cyg X-1, LMC X-1 and LMC X-3. They emphasize that the radius of any accretion disk in these three high-mass XRBs is likely to be smaller than the one derived from the orbital period since they likely transfer mass by a (possibly focused) stellar wind instead of fully developed Roche lobe overflow. In the legend, the solid horizontal line separates transient and persistent systems. (The dashed horizontal line stresses that the persistent nature of 1E 1740.7-2942 and GRS 1758-258 is unclear.) [From \cite{Coriatetal}.]}
  \label{fig:bhdim}
\end{figure}

Finally, one can test observationally if soft X-ray transients satisfy the necessary condition for instability $\dot M_{\rm tr} < \dot M_{\rm crit}(R_d)$, where $\dot M_{\rm crit}$
is the critical accretion rate for either non-irradiated or irradiated discs. In Fig. \ref{fig:bhdim} the critical accretion rates (\ref{eq:Mcritplus}) and (\ref{eq_h-irrad}) for respectively non-irradiated and irradiated disc around black holes are plotted as $\dot M(P_{\rm orb})$ relation. This relation was obtained from disc-radius -- orbital-separation relation $R_d(a)$ \cite{Paczynski77}, where (from Kepler's law) the orbital separation
$a=3.53\times 10^{10}(m_1 +m_2)^{1/3}P_{\rm hr}^{2/3}\mathrm{cm}$, where $m_i$ are the masses of the components in solar units, and $P_{\rm hr}$ the orbital period in hours.
Against these two critical lines the actual positions of the observed sources are marked. The mass transfer rate being difficult to measure, a proxy in the form of the \textsl{accumulation} rate
\begin{equation}
\label{eq:accum}
\dot M_{\rm accum}=\frac{\Delta E}{t_{\rm rec}\eta c^2}
\end{equation}
has been used. $\Delta E$ is the energy corresponding to the integrated X-ray luminosity from during an outburst and
$t_{\rm rec}$ the recurrence time of the outbursts. One can see that all low-mass-X-ray-binary (LMXB) transients are in the unstable part of the figure, as they should be if the model is correct.
One can also see that all black hole LMXBs are transient. This is not true of neutron star LMXBs. Cyg X-1 in which the stellar companion of the black hole is a massive star is observed to be stable but according to Fig. \ref{fig:bhdim} should be transient. This is not a problem because in such a system matter from the high-mass companion is not transferred by Roche-lobe overflow as in LMXBs, but lost through a stellar wind. In this case the $R_d(a)$ relation used in the plot is not valid - the discs in such systems are smaller which is marked by a left-directed arrow at the symbol marking the position of this and two other similar objects (LMC X-1 and LMC X-3).

\textsc{Additional reading}: References \cite{DHL01}, \cite{DLHC}, \cite{HMDLH}, \cite{Lasota01},  and \cite{LDK}.

\section{Black holes and advection of energy}
\label{sect:advection}

Until now, we have neglected advection terms in the energy and momentum equations
for stationary accretion flows. There two regimes of parameters where this assumption is
not valid, in both cases for the same reason: low radiative efficiency when the time for radial
motion towards the black hole is shorter than the radiative cooling time. Low density
(low accretion rate), hot, optically thin accretion flows are poor coolers and they are one
of the two configurations were advection instead of radiation is the dominant evacuation-of-energy
(``cooling") mechanism. Such optically thin flows are called ADAFs, for Advection Dominated
Accretion Flows. Also advection dominated are high-luminosity flows accreting at high rates but
they are called ``slim discs" to account for their property of not being thin but still being described
as if this were not of much importance.

We shall start with optically thin flows.

\begin{itemize}

\item ADAFs

Advection Dominated Accretion Flows' (ADAFs) is a term describing accretion
of matter with angular momentum, in which radiation efficiency is very
low. In their applications, ADAFs are supposed to describe inflows onto
compact bodies, such as black holes or neutron stars; but very hot,
optically thin flows are  bad radiators in general so that, in
principle, ADAFs are possible in other contexts. Of course in the
vicinity of black holes or neutron stars, the virial (gravitational)
temperature is $T_{\rm vir}\approx 5 \times 10^{12} (R_S/R)$ K,
so that in optically thin plasmas, at such
temperatures, both the coupling between ions and electrons and the
efficiency of radiation processes are rather feeble. In
such a situation, the thermal energy released in the flow by the
viscosity, which drives accretion by removing angular momentum, is not
going to be radiated away, but will be {\sl advected} towards the
compact body. If this compact body is a black hole, the heat will be
lost forever, so that advection, in this case, acts as sort of a `global'
cooling mechanism.  In the case of infall onto a neutron star, the
accreting matter lands on the star's surface and the (reprocessed)
advected energy will be radiated away. There, advection may act only as
a `local' cooling mechanism. (One should keep in mind that, in
general, advection may also be responsible for heating, depending on
the sign of the temperature gradient -- in some conditions, near the black hole,
advection heats up electrons in a two-temperature ADAF).

In general the role of advection in an accretion flow depends on the radiation
efficiency which in turns depends on the microscopic state of  matter and
on the absence or presence of a magnetic field. If, for a given
accretion rate, radiative cooling is not efficient, advection is
necessarily dominant, assuming that a stationary solution is possible.\\

\item Slim discs

At high accretion rates, discs around black holes become dominated by radiation
pressure in their inner regions, close to the black hole. At the same time the
opacity is dominated by electron scattering. In such discs $H/R$ is no longer
$\ll 1$. But this means that terms involving the radial velocity are no longer
negligible since $v_r \sim \alpha c_s (H/R)$. In particular, the advective term
in the energy conservation equation $v_r \partial S/\partial R$ (see Eq. \ref{eq:energy})
becomes important and finally, at super-Eddington rates, dominant. When $Q^+=Q^{\rm adv}$
the accretion flow is advection dominated and called a slim disc.

\end{itemize}

\subsection{Advection--dominated--accretion--flow toy models}

One can illustrate the fundamental properties of ADAFs and slim discs with a simple toy model.
The advection `cooling'  (per unit surface) term in the energy equation
can be written as
\begin{equation}
Q^{\rm adv} = {\dot M \over 2 \pi R^2} c_{\rm s}^2 \xi_a
\label{eq:advterm}
\end{equation}
(see Eq. \ref{xi_a}).

Using the (non-relativistic) hydrostatic equilibrium equation
\begin{equation}
{H \over R} \approx {c_{\rm s} \over v_{\rm K}}
\label{eq:hydroeq}
\end{equation}
one can write the advection term as
\begin{equation}
Q^{\rm adv} = \Upsilon \frac{\kappa_{\rm es}c}{2R}\left(\frac{\dot m}{\eta}\right)\xi_a \left({H \over R }\right)^2
\label{eq:advterm2}
\end{equation}
whereas the viscous heating term can be written as
\begin{equation}
Q^+= \Upsilon\frac{3}{8}\frac{\kappa_{\rm es}c}{R}\left(\frac{\dot m}{\eta}\right) ,             
\label{visheat}
\end{equation}
where
\begin{equation}
\label{eq:upsilon}
\Upsilon=\left(\frac{c R_S}{\kappa_{\rm es}R}\right)^2 .
\end{equation}
Since $\xi_a\sim 1$, 
\begin{equation}
\label{eq:qadv}
Q^{\rm adv}\approx Q^+\left(\frac{H}{R}\right)^2
\end{equation}
and,
as said before, for geometrically thin discs ($H/R\ll1$) the advective term $Q^{\rm adv}$ is negligible compared
to the heating term $Q^+$ and in thermal equilibrium viscous heating must be compensated by radiative cooling.
Things are different at,
very high temperatures, when $(H/R) \sim 1$. Then the advection
term is comparable to the viscous term and cannot be neglected in the equation of thermal equilibrium. In some cases
this term is larger than the radiative cooling term $Q^-$ and (most of) the heat released by viscosity is \textsl{advected} 
toward the accreting body instead of being locally radiated away as happens in geometrically thin discs. 

From Eq.(\ref{eq:amintK}) one can obtain a useful expression for the square of the relative disc height (or aspect ratio):
\begin{equation}
\left({H \over R}\right)^2 =
              {\sqrt 2 \over \kappa_{\rm es}} \left(\frac{\dot m}{\eta}\right)
                     \left(\alpha \Sigma\right)^{-1}\left(\frac{R_S}{R}\right)^{1/2}.
\label{eq:bizarre}
\end{equation}
Deriving Eq. (\ref{eq:bizarre}) we used the viscosity prescription $\nu=(2/3)\alpha c_{\rm
s}^2/\Omega_K$.

Using this equation one can write for the advective cooling
\begin{equation}
\label{eq:advterm3}
Q^{\rm adv} = \Upsilon\Omega_K \xi_a  \left(\alpha \Sigma\right)^{-1}\left(\frac{\dot m}{\eta}\right)^2.
\end{equation}
The thermal equilibrium (energy) equation is 
\begin{equation}
\label{eq:energy_eq}
 Q^+= Q^{\rm adv}+Q^- .
\end{equation}
The form of the radiative cooling term depends on the state of the accreting matter, i.e. on its temperature, density and chemical composition.
Let us consider two cases of accretion flows: 

\begin{description}
\item[--] optically thick\\
\ \ \ \  and 
\item [--] optically thin. 
\end{description}
 
For the optically thick case we will use the diffusion approximation formula
\begin{equation}
\label{eq:rad_approxQ}
Q^-= \frac{8}{3}\frac{\sigma T_c^4}{\kappa_{\rm R}\Sigma},
\end{equation}
and assume $\kappa_{\rm R}=\kappa_{\rm es}$. With the help of Eq. (\ref{eq:bizarre}) this can be brought to the form
\begin{equation}
\label{eq:slimopthick}
Q^-_{\rm thick}=8 \Upsilon \left(\frac{\kappa_{\rm es} R_S}{c}\right)^{1/2}\left(\frac{R}{R_S}\right)^2\Omega_K^{3/2} \left(\alpha \Sigma\right)^{-1/2}\left(\frac{\dot m}{\eta}\right)^{1/2} .
\end{equation}
For the optical thin case of bremsstrahlung radiation we have
\begin{equation}
\label{eq:brems}
Q^-=1.24\times 10^{21}H\rho^2T^{1/2}
\end{equation}
which using Eq. (\ref{eq:bizarre}) can be written as
\begin{equation}
\label{eq:opthin}
Q^-_{\rm thin}=3.4\times 10^{-6}\Upsilon \left(\frac{R}{R_S}\right)^2\Omega_K \alpha^{-2}\left(\alpha \Sigma\right)^2.
\end{equation}
\begin{itemize}
\item In the \textsc{optically thick} case we have therefore
\begin{eqnarray}
\label{eq:en_thick}
\xi_a \left(\frac{\dot m}{\eta}\right)^{2} +&& 0.18 \left(\frac{R}{R_S}\right)^{1/2} \left(\alpha \Sigma\right) \left(\frac{\dot m}{\eta}\right)  +\nonumber \\
&& + 2.3 \left(\frac{R}{R_S}\right)^{5/4}\left(\alpha \Sigma\right)^{1/2}\left(\frac{\dot m}{\eta}\right)^{1/2}=0
\end{eqnarray}

\item In the \textsc{optically thin} case the energy equation has the form
\begin{eqnarray}
\label{eq:en_thin}
\xi_a \left(\frac{\dot m}{\eta}\right)^{2} + && 0.18 \left(\frac{R}{R_S}\right)^{1/2} \left(\alpha \Sigma\right) \left(\frac{\dot m}{\eta}\right)  + \nonumber \\
&&+ 3\times 10^{-6}\alpha^{-2}\left(\frac{R}{R_S}\right)^2\left(\alpha \Sigma\right)^3=0
\end{eqnarray}
\end{itemize}

There are two distinct types of advection dominated accretion flows: optically thin and optically thick.
We will first deal with optically thin flows known as \textsl{ADAFs}.
\begin{figure} [!]
\centering
  \includegraphics[scale=0.4]{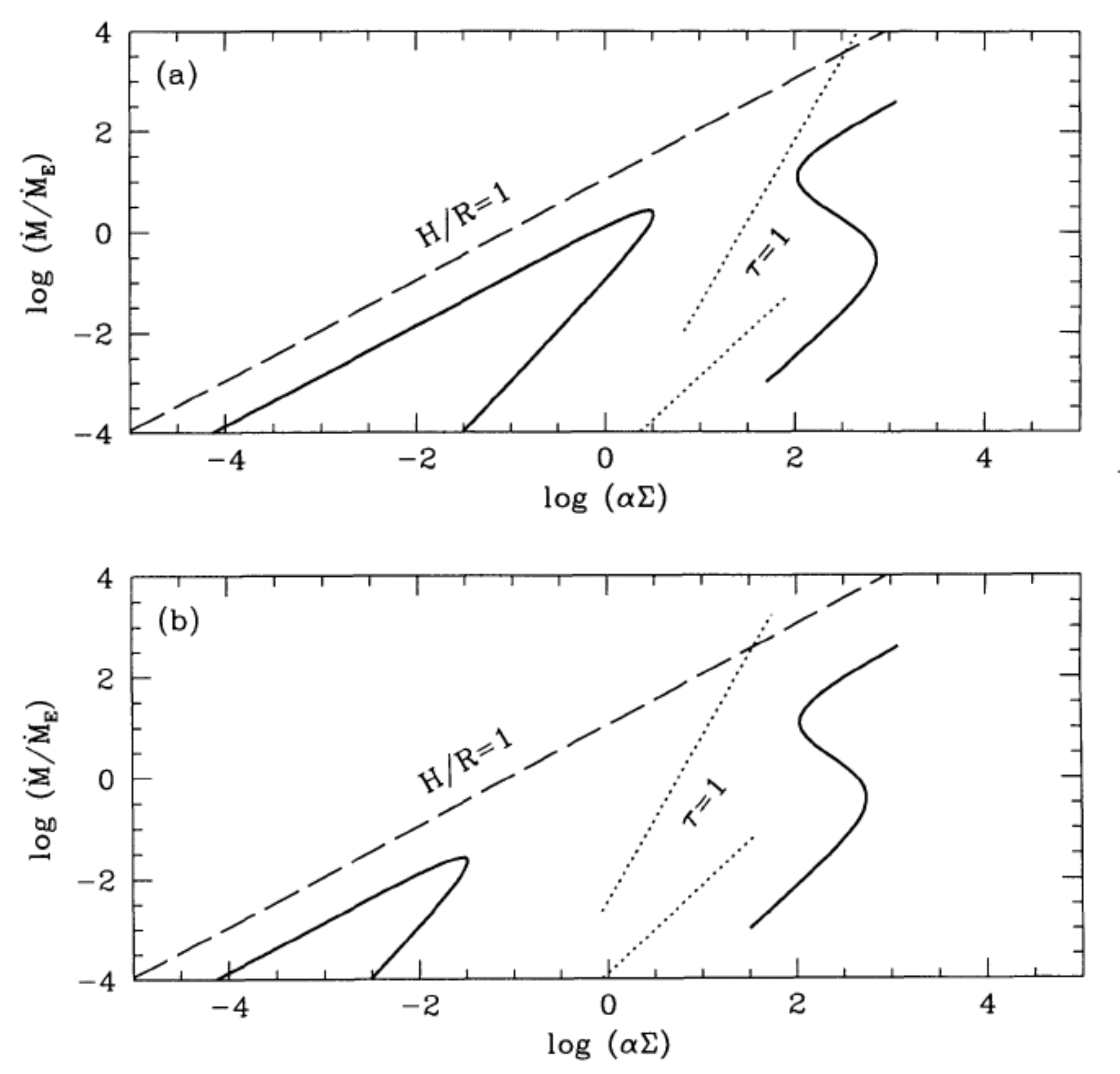}
  \caption[Thermal equilibria of accretion flows]{(a) Thermal equilibria for optically thick
  (The right solid $S$-shaped line) and optically thin (the left solid line) accretion
  flows. The upper branches represent advection-dominated solution
  (ADAFs). Flows above the dotted lines $\tau=1$ are optically thin
  -- $\tau$ is the effective optical depth calculated for
  radiation-pressure dominated (upper line) or gas-dominated
  (lower line) configurations. It is assumed that $M_{\rm
  BH}=10\msun$, $R = 5R_{\rm S}$, $\alpha=0.1$ and $\xi_a=1$. (b)
  The same for $\alpha=0.01$.[From \cite{Abramowiczetal95}]}
  \label{fig:adaf1}
\end{figure}

\subsubsection{Optically thin flows: \textsl{ADAFs}}
\label{subsub:adafthin}
For prescribed values  $\alpha$ and $\xi_a$, Eq. (\ref{eq:en_thin}) is a quadratic equation in $(\dot m/\eta)$ whose solutions in the form of $\dot m(\Sigma)$ describe thermal equilibria at a given value of $R$. Obviously, for a given $\Sigma$ this equation has at most two solutions. 
The solutions form two branches on the $\dot m(\alpha \Sigma)$ -- plane:
\begin{itemize}
\item the ADAF branch
\begin{equation}
\label{eq:adafbranch}
\dot m = 0.53\kappa_{\rm es}\,\eta \left(\frac{R}{R_S}\right)^{1/2}\xi_a^{-1} \alpha \Sigma.
\end{equation}
\noindent and\\
\item the radiatively--cooled branch
\begin{equation}
\label{eq:radcoolbranch}
\dot m = 1.9\times 10^{-5}\,\eta \left(\frac{R}{R_S}\right)^{3/2}\xi_a^{-1} \alpha^{-2}\left(\alpha\Sigma\right)^2.
\end{equation}
\end{itemize}
From Eqs. (\ref{eq:adafbranch}) and (\ref{eq:radcoolbranch}) it is clear that there exists a maximum accretion rate 
for which only one solution of Eq. (\ref{eq:en_thin}) exists. This implies the existence of a maximum accretion rate at
\begin{equation}
\dot m_{\rm max} \approx 1.7 \times 10^3 \eta \,\alpha^2 \left(\frac{R}{R_S}\right)^{1/2}.
\label{eq:mdotmax}
\end{equation}
This is where the two branches formed by thermal equilibrium solutions on the $\dot m (\alpha\Sigma)$ -- plane meet as seen on Figure \ref{fig:adaf1}.

The value of $\dot m_{\rm max}$ depends on the cooling mechanism in the accretion flow and
free-free cooling is not a realistic description of
the emission in the vicinity ($(R/R_S)\lesssim 10^3$) of a black hole. The flow there
most probably forms a two-temperature plasma. In such a case
$\dot m_{\rm max}\approx 10 \alpha^2$ with
almost no dependence on radius. For larger radii $\dot m_{\rm max}$
decreases with radius.

\subsubsection{Optically thick flows: \textsl{slim discs}}
\label{subsub:adafthick}

Since the first two terms in Eq. (\ref{eq:en_thick}) are the same as in (Eq. \ref{eq:en_thin}), the high $\dot m$, advection dominated solution is
the same as in the optically thin case but now represents the
\begin{itemize}
\item Slim disc branch
\begin{equation}
\tag{\ref{eq:adafbranch}}
\dot m = 0.53\, \kappa_{\rm es}\,\eta \left(\frac{R}{R_S}\right)^{1/2}\xi_a^{-1} \alpha \Sigma.
\end{equation}

Now, the full equation (\ref{eq:en_thick}) is a cubic equation in $\dot m^{1/2}$ and on the $\dot m(\alpha \Sigma)$ plane its solution forms the
two upper branches of the \bS-curve shown in  Fig. \ref{fig:adaf1}. The uppermost branch corresponds to slim discs while the branch with negative slope 
represents the Shakura-Sunayev solution in the regime \textsl{a.} (see Sect. \ref{subsec:SS}), i.e. \\
\item a radiatively cooled, radiation-pressure dominated accretion disc
\begin{equation}
\label{eq:radpdom}
\dot m =160\,\kappa^{-1}_{\rm es}\, \eta \left(\frac{R}{R_S}\right)^{3/2}\left(\alpha \Sigma\right)^{-1}
\end{equation}
\end{itemize}

\subsubsection{Thermal instability of radiation--pressure dominated discs}

Radiation--pressure dominated ($P=P_{\rm rad}$) accretion discs are thermally unstable when
opacity is due to electron scattering on electrons. Indeed
\begin{equation}
\frac{d\ln T_{\rm eff}^4}{ d\ln T_{\rm c}} = 4
\label{eq:cooles}
\end{equation}
because $\kappa_R=\kappa_{\rm es}=const.$, while
in a radiation pressure dominated disc $Q^+\sim \nu\Sigma \sim HT^4\sim T^8/\Sigma$
so
\begin{equation}
\frac{d\ln Q^+}{d\ln T_{\rm c}} = 8 \, > \, \frac{d\ln T_{\rm eff}^4}{ d\ln T_{\rm c}}
\label{eq:heates}
\end{equation}
and the disc is thermally unstable. This solution is represented by the middle branch with negative slope (see Eq. \ref{eq:radpdom}) in Fig. \ref{fig:adaf1}.
The presence of this instability in the model is one of the unsolved problems of the accretion disc theory because it contradicts observations which do not
show any unstable behaviour in the range of luminosities where discs should be in the radiative pressure and electron-scattering opacity domination regime.

\subsubsection{Slim discs and super-Eddington accretion}

From Eqs. (\ref{eq:bizarre}) and (\ref{eq:radpdom}) one obtains for the disc aspect ratio
\begin{equation}
\label{eq:htorradpress}
\frac{H}{R}=0.11\left( \frac{\dot m}{\eta}\right)\frac{R_S}{R}
\end{equation}
which shows that the height of a radiation dominated disc is constant with radius and proportional to the accretion rates.

But this means that with increasing $\dot m$ advection becomes more and more important (see e.g. Eq. \ref{eq:qadv}) and for
\begin{equation}
\label{eq:adveqplus}
\frac{\dot m}{\eta}\approx 9.2 \frac{R}{R_S}
\end{equation}
advection will take over radiation as the dominant cooling mechanism and the solution will represent a slim disc.
Equation (\ref{eq:adveqplus}) can be also interpreted as giving the \textsl{transition radius} between radiatively and
advectively cooled disc for a given accretion rate $\dot m$:
\begin{equation}
\label{eq:advectrans}
\frac{R_{\rm trans}}{R_S}\approx \frac{0.1}{\eta}{\dot m}
\end{equation}
Another radius of interest is the \textsl{trapping radius} at which the photon diffusion (escape)  time $H\tau/c$ is equal to the
viscous infall time $R/v_r$
\begin{equation}
\label{eq:trapp}
R_{\rm trapp}= \frac{H\tau\,v_r}{c}= \frac{H\kappa \Sigma}{c}\frac{\dot M}{2\pi R \Sigma}=\frac{H}{R}\left(\frac{\dot m}{\eta}\right)R_S.
\end{equation}
Notice that both $R_{\rm trans}$ and $R_{\rm trapp}$ are proportional to the accretion rate.

In an advection dominated disc the aspect ration $H/R$ is independent of the accretion rate:
\begin{equation}
\label{eq:htorslim}
\frac{H}{R}=0.86\, \xi_a\left(\frac{R}{R_S}\right)^{1/4},
\end{equation}
therefore contrary to radiatively cooled discs, slim disc do not puff up with increasing accretion rate.

Putting (\ref{eq:htorslim}) into Eq. (\ref{eq:trapp}) one obtains
\begin{equation}
\label{eq:trapp2}
\frac{R_{\rm trapp}}{R} = 0.86\,\xi_a^{-1/2}\left(\frac{R}{R_S}\right)^{1/4}\left(\frac{\dot m}{\eta}\right).
\end{equation}
Radiation inside the trapping radius is unable to stop accretion and since $R_{\rm trapp}\sim \dot m$ there is no
limit on the accretion rate onto a black hole.

The luminosity of the toy-model slim disc can be calculated from Eqs. (\ref{eq:slimopthick}) and (\ref{eq:adafbranch})
giving
\begin{equation}
\label{eq:Qminusslim}
Q^-=\sigma T^4_{\rm eff}=\frac{0.1}{\xi_a}\frac{L_{\rm Edd}}{R^2},
\end{equation}
which implies $T_{\rm eff}\sim 1/R^{1/2}$.
The luminosity of the slim--disc part of the accretion flow is then
\begin{equation}
\label{eq:luminslim}
L_{\rm slim}=2\int_{R_{\rm in}}^{R_{\rm trans}}\sigma T^4_{\rm eff}2\pi RdR=\frac{0.8}{\xi_a}L_{\rm Edd}\cdot\ln\frac{R_{\rm trans}}{R_{\rm in}}\approx L_{\rm Edd}\ln\dot m,
\end{equation}
where we used Eq. (\ref{eq:advectrans}).

Therefore the total disk luminosity
\begin{eqnarray}
\label{eq:total}
L_{\rm total}&=& L_{\rm thin} + L_{\rm slim}=\\
&& 4\pi\left(\int_{R_{\rm in}}^{R_{\rm trans}}\sigma T^4_{\rm eff} RdR+ \int_{R_{\rm trans}}^{R_{\infty}}\sigma T^4_{\rm eff}RdR\right)\approx L_{\rm Edd}(1 + \ln\dot m), \nonumber 
\end{eqnarray}
where $L_{\rm thin}$ is the luminosity of the radiation-cooled disc for which Eq. (\ref{eq:teff}) applies.

It is easy to see that the same luminosity formula $L\approx L_{\rm Edd}(1 + \ln\dot m)$ is obtained when one assumes mass--loss from the disc resulting in a variable (with radius) accretion rate: $\dot M \sim R$.

At very high accretion rates the disc emission will be also strongly beamed by the flow geometry so that and observer situated in the beam of the
emitting system will infer a luminosity
\begin{equation}
\label{eq:beam}
L_{\rm sph}=\frac{1}{b}L_{\rm Edd}(1 + \ln\dot m),
\end{equation}
where $b$ is the beaming factor (see \cite{King09} for a derivation of $b$ in the case of Ultra-Luminous X-ray sources).

Numerical simulations do not seem to correspond to this analytical solutions (see e.g. \cite{Jiang14}, \cite{SadowskiNarayan15} and \cite{Sadowskietal14} but they also disagree between themselves. The reasons for these contradictions are worth investigating.

\textsc{Additional reading:} References \cite{Abramowicz05}, \cite{Abramowiczetal88}, \cite{Abramowiczetal95}, \cite{LKD}, \cite{NarayanYi94}, \cite{Sadowski09}, \cite{Sadowski11}, and \cite{YuanNarayan14}.


\section{Accretion discs in Kerr spacetime.}
\label{sec:kerr}

In this section we will present and discuss the set equations whose solutions represent $\alpha$--accretion discs
in the Kerr metric. This section is based on references \cite{Abramowiczetal96}, \cite{Lasota94} and \cite{Sadowski09} and
to be understood requires some basic knowledge of Einstein's General Relativity.

\subsection{Kerr black holes}

The components $g_{ij}$ of the metric tensor with respect to the
coordinates $(t,x^\alpha)$ are expressible in terms of the lapse
$N$, the components $\beta^\alpha$ of the shift vector and the
components $\gamma_{\alpha \beta}$ of the spatial metric:
\begin{equation}
\label{fol} g_{ij} \, dx^i\, dx^j
    = - N^2 dt^2 + \gamma_{\alpha \beta} (dx^\alpha + \beta^\alpha dt)
        (dx^\beta + \beta^\beta dt) ,
\end{equation}
which is a modern way of writing the metric.
  
\textsc{Remark 3.}
In this section only I will use conventions different from those used in other parts of the Chapter. First, I will use the so-called \textsl{geometrical  units} that are linked to the physical units for length, time and mass by
\begin{eqnarray}
\rm length~in~physical~units &=& \rm ~length~in~geometrical~units, \nonumber\\
\rm time~in~physical~units &=& \rm ~\frac{1}{c}~length~in~geometrical~units, \nonumber\\
\rm mass~in~physical~units &=& \rm ~\frac{c^2}{G}~length~in~geometrical~units. \nonumber\\
\end{eqnarray}
Second, the radial coordinate will be called ``$r$" and not ``$R$". This should not confuse the reader since $R$ is used only
in the \textsl{non-relativistic} context where it denotes a radial coordinate and a radial distance, while in the relativistic context it only
a coordinate.

\subsubsection{General structure, Boyer-Lindquist coordinates}

The Kerr metric in the Boyer-Lindquist (spherical) coordinates $t, r, \theta, \varphi$
corresponds to:
\begin{eqnarray}
N=\frac{\varsigma}{\sqrt{A\Delta}},\ \ \ \beta^r=\beta^\theta=0,\ \ \ \beta^\varphi= - \omega, &&  \\
g_{rr}=\frac{\varsigma^2}{\Delta}, \ \ \
g_{\theta\theta}=\varsigma^2,\ \ \
g_{\varphi\varphi}=\frac{A^2}{\varsigma^2}\sin^2 \theta &&
\label{Ng}
\end{eqnarray}
with
\begin{equation}
\varsigma= r^2 + a^2 \cos^2 \theta,\ \ \  \Delta = r^2 - 2 Mr + a^2,
\label{varsigma}
\end{equation}
\begin{equation}
A = \left(r^2 + a^2\right)^2 - \Delta a^2\sin^2 \theta, \ \ \ \omega
= \frac{2Jr}{A}={2Mar\over A},
\label{kerrmet}
\end{equation}

\noindent where $M$ is the mass and $a=J/M$ is the angular momentum per
unit mass. In applications one often uses the dimensionless
``angular-momentum" parameter $a_*=a/M$.

Therefore in BL coordinates the Kerr metric takes the form of
\begin{equation}
ds^2 = - {{\varsigma^2\Delta}\over A}dt^2 + {A \sin^2 \theta \over
\varsigma^2}\left (d\varphi - \omega dt\right)^2 + {\varsigma^2\over
\Delta}dr^2 + \varsigma^2 d\theta^2.
\label{metr}
\end{equation}

The time (stationarity) and axial symmetries of the metric are expressed by two Killing vectors
\begin{equation}
 {\eta^i} = \delta^i_{~(t)}, ~~{\xi^i}= \delta^i_{~(\varphi)},
\label{kill}
\end{equation}
where $\delta^i_{~(k)}$ is the Kronecker delta.\\

\textsc{Remark 4.} Using Killing vectors (\ref{kill}) one can define some useful
scalar functions: the angular velocity of the dragging of inertial
frames $\omega$, the gravitational potential $\Phi$, and the
gyration radius ${\mathfrak R}$,

\begin{equation}
 \omega = - \frac{\vec{\eta} \cdot \vec{\xi}}{\vec{\xi} \cdot \vec{\xi}},~~ e^{-2\Phi}
 =
\omega^2 {\vec{\xi}} \cdot \vec{\xi}- {\vec{\eta}\cdot \vec{\eta}},~~ {\mathfrak
R}^2 = - \frac{\vec{\xi}\cdot \vec{\xi}}{\vec{\eta}\cdot \vec{\eta}}.
\label{functions}
\end{equation}

\noindent In the Boyer-Lindquist coordinates the scalar products of the
Killing vectors are simply given by the components of the metric,

\begin{equation}
{\vec{\eta}\cdot \vec{\eta}} = g_{tt},~~ {\vec{\eta} \cdot \vec{\xi}}= g_{t\varphi},
~~ {\vec{\xi} \cdot \vec{\xi}} = g_{\varphi \varphi},
\label{kill2}
\end{equation}

\noindent and therefore quantities defined in Eq. (\ref{functions}) can be explicitly
written down in terms of the Boyer-Lindquist coordinates as:

\begin{equation}
{\mathfrak R}^2 = \frac{A^2}{{r^4 \Delta}},~~e^{-2\Phi} =
\frac{{r^2\Delta}}{A}.
\end{equation}

\begin{itemize}

\item \textbf{The horizon}

\end{itemize}

\noindent The black hole surface (event horizon) is at

\begin{equation}
r_H= M + \sqrt{M^2 - a^2}.
\end{equation}

Therefore a horizon exists for ${a}_* \leq 1$ only. At the horizon the
angular velocity of the dragging of inertial frame is equal to

\begin{equation}
\omega_{H}= \Omega_{\rm H} = \frac{a}{2Mr_H},
\end{equation}
where
$\Omega_{\rm H}$ is the angular velocity of the horizon, i.e. the
angular velocity of the horizon-forming light-rays with respect to
infinity. The horizon rotates.

\noindent The area of the horizon is given by

\begin{equation}
S= 8 \pi M r_H= 8\pi M \sqrt{M^2 - a^2}.
\end{equation}

The extreme (maximally rotating) black hole corresponds to
\begin{equation}
a=M.
\end{equation}
For $a>M$ the Kerr solution represents a \textsl{naked singularity}.
Such singularities would be a great embarrassment not only because of their
visibility but also because the solution of Einstein equation in which they appear violate
causality by containing closed time-like lines.
The conjecture that no naked singularity is formed through collapse
of real bodies is called the \emph{cosmic censorship hypothesis} (Roger Penrose).

\textsc{Remark 5.}
\textbf{Rotation of astrophysical bodies}\\
Since this is a lecture in astrophysics let us leave for a moment
the geometrical units. They are great for calculations but usually useless
for comparing their results with observations. In the physical units
\begin{equation}
r_H = \frac{GM}{c^2}+ \left[\left(\frac{GM}{c^2}\right)^2 -
\left(\frac{J}{Mc}\right)^2 \right]^{1/2}.
\end{equation}
and therefore the
maximum angular momentum of a black hole is
\begin{equation} J_{\rm max}=
\frac{GM^2}{c} = 8.9 \times 10^{48} \left(\frac{M}{\msun}\right)^2
\rm g\ cm^2\ s^{-1}.
\end{equation}
This is slightly more than the angular
momentum of the Sun ($J_{\odot}= 1.63 \times 10^{48} \rm g\ cm^2\
s^{-1}$, ${a}^{\odot}_*= 0.185$): the gain in velocity is almost fully
compensated by the loss in radius.

For a millisecond pulsar which is a neutron star with a mass of $\sim 1.4 \msun$
and radius $\sim 10 \rm km$ the angular momentum is
\begin{equation}
J_{\rm NS}= I_{\rm NS}\Omega_S
\approx 8.6\times 10^{48}
\left(\frac{\alpha(x)}{0.489}\right)
\left(\frac{M_{\rm NS}}{1.4\,\msun}\right)
\left(\frac{R_{\rm NS}}{10\, \rm km}\right)^2
\left(\frac{P_S}{\rm 1\, ms}\right)^{-1}
{\rm g\ cm^2}
\end{equation}
where $I_{\rm NS}\approx \alpha(x) M_{\rm NS}R_{\rm NS}^2$ is the moment of
inertia and $x=(M_{\rm NS}/\msun)(km/R_{\rm NS})$ the
compactness parameter. For the most compact neutron star $x\leq 0.24$
and $\alpha(x)\lesssim 0.489$.
Therefore for neutron stars that rotate at millisecond periods
\begin{equation}
a^{\rm NS}_* \approx 0.5 \left(\frac{\alpha(x)}{0.489}\right)
\left(\frac{M_{\rm NS}}{1.4\,\msun}\right)^{-1}
\left(\frac{R_{\rm NS}}{10\, \rm km}\right)^2
\left(\frac{P_S}{\rm 1\, ms}\right)^{-1}.
\end{equation}

\noindent By definition\\
\begin{itemize}

\item  the specific (per unit rest-mass) energy is

\end{itemize}

\begin{equation}
\mathfrak{E} := - {\vec{\eta} \cdot \vec{u}},
\end{equation}

\begin{itemize}
\item the specific (per unit rest-mass) angular momentum
\end{itemize}
\begin{equation}
\mathfrak{L}:= \vec{\xi} \cdot \vec{u}
\end{equation}
and\\
\begin{itemize}
\item the specific (per unit mass-energy) angular momentum (also called \textsl{geometrical} specific angular momentum)
\end{itemize}
\begin{equation}
\mathfrak{J}:= -\frac{\mathfrak{L}}{\mathfrak{E}} = - \frac{ \vec{\xi} \cdot \vec{u}}{{\vec{\eta} \cdot \vec{u}}}
\end{equation}

\subsection{Privileged observers}

Let us consider observers privileged by the symmetries of the Kerr
spacetime. The results below apply to any spacetime with the same
symmetries, e.g. the spacetime of a stationary, rotating star. The
four-velocity of a privileged observer is the linear combination of
the two Killing vectors:
\begin{equation}
{\vec{u}}= Z\left(\vec{\eta} + \Omega_{\rm obs} \vec{\xi}\right)
\end{equation}
where the redshift
factor $Z$ is (from the normalization ${\vec{u}}\cdot{\vec{u}}=1$)
\begin{equation}
Z^{-2}=\vec{\eta} \cdot \vec{\eta} + 2\Omega_{\rm obs} \vec{\eta} \cdot \vec{\xi} + \Omega_{\rm
obs}^2 \vec{\xi }\cdot \vec{\xi}
\end{equation}

Since for $a\neq 0$ the Kerr spacetime is stationary but not static, i.e;
the timelike Killing vector $\eta$ is not orthogonal to the space-like
surfaces $t = $const. In such a spacetime "non-rotation" is not uniquely
defined.

{\sl Stationary} observers are
immobile with respect to infinity; their four-velocities are defined
as
\begin{equation}
u^i_{\rm stat}= \left(\eta \eta\right)^{-1/2}\eta^i
\end{equation}
but are locally rotating: ${\cal L}_{\rm stat}=\xi_iu^i_{\rm stat}\neq 0$.

\noindent The four-velocity of a locally non-rotating observer is a unit timelike
vector orthogonal to the space-like surfaces $t =
$const.:
\begin{equation}
 u^i_{\rm ZAMO} = e^{\Phi}\left (\eta^i + \omega \xi^i\right ),
\label{unit}
\end{equation}
defines four-velocity of the local inertial observer or ZAMO, i.e.
Zero Angular Momentum Observers since
\begin{center}
${\cal L}_{\rm ZAMO}=\xi_i u^i_{\rm ZAMO}=0$.
\end{center}

Finally, in presence of matter forming a stationary and axisymmetric
configuration, there are privileged observers \emph{comoving} with matter.

\begin{table} [ht!]
\begin{center}
\caption{Summary of properties of privileged observers}
\begin{tabular}{|l|l|c|}
\hline
 & &  \\
~~Observer &~~Four-velocity &Angular velocity with respect \\
&~~~~~~~~~~~&to stationary observers\\

\hline
& &  \\ ~~Stationary &$\ \ \vec{u}=\left(\vec{\eta} \cdot \vec{\eta}\right)^{-1/2} \vec{\eta}$ &$\Omega_{\rm stat}=0\ \ $ \\
&~~~~~~~~~~~&~~~~~~~~~~~~\\~~ZAMO (LNR) &\ \ ${\vec{u}}=e^{-\Phi}\left({\vec{\eta}} + \omega{\vec{\xi}}\right)\ \ $ &$\Omega_{\rm ZAMO}=\omega$  \\
&~~~~~~~~~~~&~~~~~~~~~~~~\\~~Comoving (with matter) ~~& $\ \ {\vec{u}}=Z\left({\vec{\eta}} + \Omega{\vec{\xi}}\right)\ \ $ &$\Omega_{\rm com}=\Omega$ \\
& &  \\
\hline
\end{tabular}
\end{center}
\label{table-obs}
\end{table}

\subsection{The ergosphere}

For ZAMOs $\Omega=\omega$ but for  stationary observers ${ \Omega -
\omega}= - \omega$. Therefore ZAMOs rotate with respect to infinity
(but are {\sl locally} non-rotating). They may exist down to the
black hole horizon, where they become null: $\vec{u}_{\rm ZAMO}\
\cdot \vec{u}_{\rm ZAMO}=0$.

Stationary observers immobile with respect to infinity but rotating with angular velocity $-\omega$ with
respect to ZAMOs can exist (their four-velocity must be timelike, $\vec{\eta} \cdot \vec{\eta} <0$)
only outside the \emph{stationarity limit} whose radius is defined by $\vec{\eta} \cdot \vec{\eta} =0$:
\begin{equation}
r_{\rm er}(\theta)= M + \sqrt{M^2 - a^2\cos^2 \theta}.
\end{equation}
The stationarity limit is called the \emph{ergosphere}.

\subsection{Equatorial plane}
We will discuss now orbits in the equatorial plane, where they have
the axial symmetry.
We are introducing the cylindrical vertical coordinate $z = \cos \theta$ is defined
very close to the equatorial plane, $z = 0$. The metric of the Kerr
black hole in the equatorial plane, accurate up to the $(z/r)^0$ is

\begin{equation}
ds^2 = - \frac{{r^2\Delta}}{A}dt^2 + \frac{A}{r^2}\left
(d\varphi - \omega dt\right)^2 + {r^2}\frac{\Delta}dr^2 + dz^2,
\label{metreq}
\end{equation}
where now
\begin{equation} \Delta = r^2 - 2 Mr + a^2,~~ A=\left(r^2 + a^2\right)^2 -
\Delta a^2, ~~\omega = \frac{2Mar}{A},
\end{equation}

\noindent or simpler
\begin{equation}
ds^2 = - \left(1 - \frac{2M}{r}\right)dt^2 -2\omega dtd\varphi + \frac{A}{r^2}d\varphi^2
+ \frac{r^2}{\Delta}dr^2 + dz^2 .
\end{equation}

\subsubsection{Orbits in the equatorial plane}
\label{orbits}

\noindent The four velocity of matter $u^i$ has components $u^t$,
$u^{\varphi}$, $u^r$,
\begin{equation}
 u^i = u^t \delta^i_{~(t)} + u^{\varphi} \delta^i_{~(\varphi)}
       + u^r \delta^i_{~(r)}.
\label{fourv}
\end{equation}
The angular frequency $\Omega$ with respect to a stationary observer,
and the angular frequency ${\tilde \Omega}$ with respect to a local
inertial observer are respectively defined by
\begin{equation}
 \Omega = \frac{u^{\varphi}}{u^t},~~ {\tilde \Omega} =
\Omega - \omega, \label{omega}
\end{equation}

\noindent The angular frequencies of the corotating (+) and
counterrotating (--) \emph{Keplerian orbits} are
\begin{equation}
\Omega_K^{\pm}=\pm\frac{M^{1/2}}{r^{3/2}\pm aM^{1/2}},
\label{omegaK}
\end{equation}

\noindent the specific energy is
\begin{equation}
{\mathfrak E}_K^{\pm}
= \frac{r^2 - 2Mr \pm a(Mr)^{1/2}}
{r\left(r^2 - 3Mr \pm 2a(Mr)^{1/2}\right)^{1/2}}    
\label{ekepler}         
\end{equation}

\noindent and the specific angular momentum is given by
\begin{equation}
{\mathfrak L}_K^{\pm}
= \pm  \frac{\left(Mr\right)^{1/2}\left(r^2 \mp 2a(Mr)^{1/2} + a^2\right)}
{r\left(r^2 - 3Mr \pm 2a(Mr)^{1/2}\right)^{1/2}},
\label{lkepler}
\end{equation}
or
\begin{equation}
\label{eq:geoam}
{\mathfrak J}_K=  \pm  \frac{(Mr)^{1/2}\left(r^2 \mp 2a(Mr)^{1/2} + a^2\right)}{r^2 - 2Mr \pm a(Mr)^{1/2}}.
\end{equation}
Both ${\mathfrak J}$ and ${\mathfrak L}$ have a minimum at the last stable orbit, more often called ISCO
(Innermost Stable Circular Orbit).\\

Because of the rotation of space there is no direct relation between angular momentum and angular frequency
but
\begin{equation}
\label{}
{\mathfrak J}=\frac{\vec{\eta} \cdot \vec{\xi} + \Omega\vec{\xi} \cdot \vec{\xi}}{\vec{\eta}\cdot \vec{\eta} + \Omega\vec{\eta} \cdot \vec{\xi}}
=
\frac{{\mathfrak R}^2\Omega - \omega}{\Omega\omega -1}.
\end{equation}

For the Schwarzschild solution ($a=\omega=0$)
\begin{equation}
\label{eq:kepler1}
{\mathfrak J}_K= {\mathfrak R}^2\Omega_K,
\end{equation}
so a Newtonian-like relation (justifying the name ``gyration radius" for ${\mathfrak R}$) between angular frequency and angular momentum exists
for ${\mathfrak J}$. No such relation exists for ${\mathfrak L}$.\\

\noindent {\bf ISCO}\\

The minimum of the Keplerian angular momentum corresponding to the \textsl{innermost stable circular orbit} (ISCO) is located at
\begin{eqnarray}
r_{\rm ISCO}^{\pm}& = & M \{ 3 + Z_2 \mp [(3 - Z_1)(3 + Z_1 + 2 Z_2)]^{1/2}\},
\nonumber \\
Z_1 & =& 1 +\left(1 - a^2/M^2 \right)^{1/3}\left[(1 + a/M)^{1/3} +
         (1 - a/M)^{1/3} \right],
\nonumber \\
Z_2 & =& \left(3 a^2/M^2 + Z_1^2 \right)^{1/2}.
\end{eqnarray}
\begin{figure} [!]
\includegraphics[angle=0,scale=0.23]{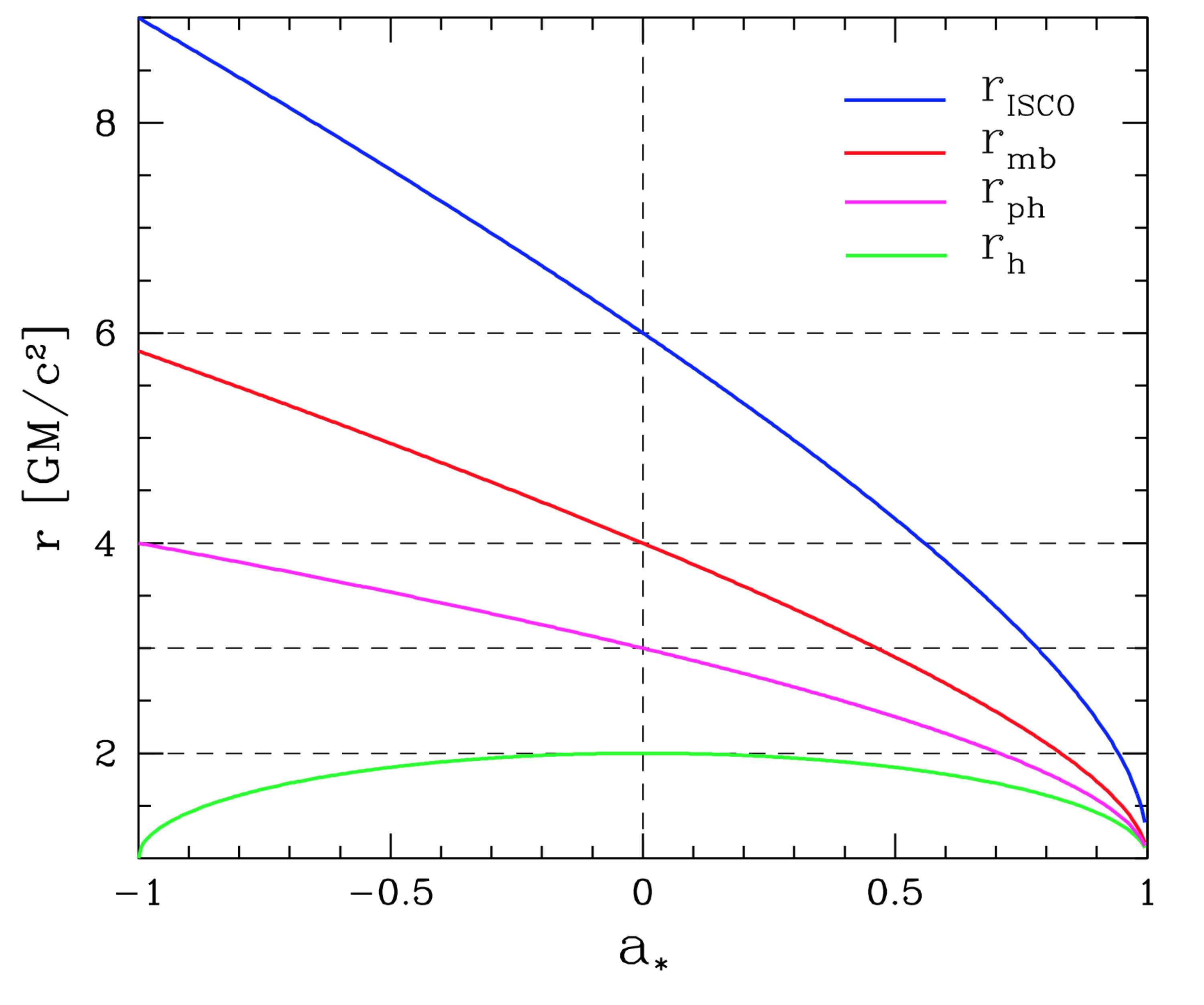}
\caption{Radii of characteristic orbits in the Kerr metric as a function of $a_*=a/M$. The innermost stable circular orbit: $r_{\rm ISCO}$, the marginally bound orbit $r_{\rm IBCO}$ (marked $r_{\rm mb}$), the photon orbit: $r_{\rm ph}$ and the black hole horizon: $r_{H}$ (marked $r_{\rm h}$). (Courtesy of A. S{\c a}dowski.)}
\label{fig:freq0}
\end{figure}

\noindent \textbf{Binding energy}\\

The binding energy 
\begin{equation}
\label{eq:bindingE}
\mathcal{E}_{\rm bind}=1 - {\mathfrak E}_K
\end{equation}
at the ISCO  is
\begin{itemize}
\item $1 - \sqrt{8/9} \approx 0.06$ for $a=0$ 
\item $1 - \sqrt{1/3} \approx 0.42$ for $a=1$. 
\end{itemize}
This corresponds to the efficiencies
of accretion in a geometrically thin (quasi-Keplerian) disc around a black hole.

For a Schwarzschild black hole the frequency associated with the ISCO at $r_{\rm ISCO}=6M$ is 
\begin{equation}
\nu_K(r_{\rm ISCO}) = 2197\,\left(\frac{M}{\msun}\right)^{-1}\ \rm Hz.
\end{equation}

\noindent {\bf IBCO}\\

The binding energy of a Keplerian orbit $1 - {\mathfrak E}=0$ at the marginally bound
orbit (or IBCO: Innermost Bound Circular Orbit)
\begin{equation}
r_{\rm IBCO}^{\pm}= 2M \mp a + 2\sqrt{M^2 \mp aM}.
\end{equation}
For a non-rotating black-hole $r_{\rm IBCO}=4M$ and the frequency associated with the IBCO is
\begin{equation}
\nu_K(r_{\rm IBCO}) = 4037\,\left(\frac{M}{\msun}\right)^{-1}\ \rm Hz.
\end{equation}

\noindent {\bf ICO ({\sl Circular photon orbit})}\\

The Innermost Circular Orbit (ICO), i.e. the circular photon orbit is at
\begin{equation}
r_{\rm ph}^{\pm}= 2M \left(1 + \cos\left[\frac{2}{3}\cos^{-1}\left(\mp \frac{a}{M}\right)\right]\right).
\end{equation}
For a non-rotating black hole $r_{\rm ph}=3M$.

\subsubsection{Epicyclic frequencies}

We will consider now consider a perturbed orbital motion in, and slightly off the equatorial plane. 
In the Newtonian case the angular frequency of such motions must be equal to the Keplerian frequency 
$\Omega_K$ since in there is only one characteristic scale defined by the gravitational constant $G$. 
In General Relativity the presence of two constants $G$ and $c$ imply that the epicyclic frequency does
not have to be equal to $\Omega_K$.

The four--velocity for the perturbed circular motion can be written as
\begin{equation}
u^i = \left(1,\tilde u^r,\tilde u^\theta,\Omega_K + \tilde u^\varphi\right),
\end{equation}
where $\tilde u^{\alpha}$ are the velocity perturbations.\\
\begin{itemize}
\item For perturbations in the equatorial plane the equation of motion is
\begin{equation}
\left( \der{^2}{t^2} + \kappa^2 \right)
\left(\begin{array}{c}{\tilde u^r}\\
{\tilde u^\varphi}\\
\end{array}\right)
=0,
\label{eomep}
\end{equation}
\noindent where
\begin{equation}
\kappa^2=\Omega_K^2 \frac{r^2-6Mr \pm 8aM^{1/2}r^{1/2} - 3a^2}{r^2}
\end{equation}
is the (equatorial) \textsl{epicyclic frequency}.
In the Schwarzschild case $a=0$ this is $\kappa^2=\Omega_K^2 (1-6M/r)$ and vanishes
at ISCO. In the Newtonian limit the epicyclic frequency equal to the Keplerian frequency $\kappa=\Omega_K$.\\

\item \noindent For vertical perturbations the equation is
\begin{equation}
\left( \der{}{t} + \Omega_K \der{}{\varphi}\right) \tilde u^{\theta}= - \Omega_\bot \delta \theta,
\end{equation}
where the vertical epicyclic (angular) frequency is given by
\begin{equation}
\Omega_\bot^2= \Omega_K^2\frac{r^2 - {4a}M^{1/2}{r^{1/2}} - {3a^2M^2}}{r^2}
\end{equation}
In the Schwarzschild case ($a=0$) the vertical epicyclic frequency is equal to the Keplerian angular frequency $\Omega_K$, which
is to be expected from the spherical symmetry of this solution.

The angular velocity $\Omega_\bot$ appears also in the equation of vertical equilibrium of a (quasi)Keplerian disc which
will be discussed later (sect. \ref{sub:verteq}).
Here let us just notice that Eq.(\ref{eq:vertrel}) can be written as
\begin{equation}
\der{p}{z}=- \rho e^{2\Phi} \Omega_\bot^2 z.
\end{equation}
\end{itemize}

All these characteristic frequencies can be put into the form
\begin{equation}
\Omega = f\left(x,a_*\right)\frac{1}{M},
\end{equation}
where $x=r/M$. For all relativistic frequencies $x=x(a_*)$ and therefore
they can be written as
\begin{equation}
\Omega = F\left(a_*\right)\frac{1}{M}.
\label{overm}
\end{equation}

\textsc{Additional reading}: Reference \cite{AK2004}.

\section{Accretion flows in the Kerr spacetime}
\label{sec:accretion_Kerr}

\subsection{Kinematic relations}

In the reference frame of the local inertial (non-rotating) observer
the four velocity takes the form,
\begin{equation}
 u^i = \gamma \left (u_{\rm ZAMO}^i + v^{(\varphi)}\tau^i_{~(\varphi)} +
v^{(r)}\tau^i_{~(r)}\right). \label{fourv2}
\end{equation}
The vectors $\tau^i_{~(\varphi)}$ and $\tau^i_{~(r)}$ are the unit
vectors in the coordinate directions $\varphi$ and $r$. The Lorentz
gamma factor $\gamma$ equals,
\begin{equation}
 \gamma = {1\over {\sqrt{1 - \left(v^{(\varphi)}\right)^2 -
\left(v^{(r)}\right)^2}}}. \label{gamma}
\end{equation}
The relation between the Boyer-Lindquist and the physical velocity
component in the azimuthal direction is,
\begin{equation}
  v^{(\varphi)} = {\tilde R}{\tilde \Omega},
\end{equation}
which justifies the name of $\tilde R$ -- gyration radius. It is convenient to use the (rescaled) radial
velocity component $V$ defined by the formula,
\begin{equation}
  {V \over {\sqrt {1 - V^2}}} = \gamma v^{(r)} = u^r g_{rr}^{1/2}.
\label{V}
\end{equation}
The Lorentz gamma factor may then be written as,
\begin{equation}
\gamma^2=\left({1\over1-{\tilde \Omega}^2{\tilde R}^2}\right)
\left({1\over1-V^2}\right), \label{gamma2}
\end{equation}
which allows writing a simple expression for $V$ in terms of the
velocity components measured in the frame of the local inertial
observer,
\begin{equation}
 V = {{v^{(r)}}\over{\sqrt{1 - \left(v^{(\varphi)}\right)^2}}} =
{{v^{(r)}}\over{\sqrt{1 - {\tilde R}^2{\tilde \Omega}^2}}}.
\label{simpleV}
\end{equation}
Thus, $V$ is the radial velocity of the fluid as measured by an
observer corotating with the fluid at fixed $r$.

Although a different quantity could have been chosen as the
definition of the ``radial velocity'', only $V$ has directly three very
convenient properties, all guaranteed by its definition:
\begin{itemize}
\item (i) everywhere in the flow $|V| \le 1$,
\item (ii) on the horizon $|V| = 1$,
\item (iii) at the sonic point $|V| \approx c_s$,
\end{itemize}
where $c_s$ is the local sound speed.

To see that property (i) holds, let us define
\begin{equation}
{\tilde V}^2 = u^r u_r = u^r u^r g_{rr} \ge 0.
\end{equation}
Then, one has
\begin{equation}
V^2 = {\tilde V}^2/(1 + {\tilde V}^2) \le 1.
\end{equation}

Writing  $V = \sqrt{r^2 u^r u^r/(r^2 u^r u^r + \Delta)}$ demonstrates
property (ii) since $|V|=1$ independent of the value of $r^2 u^r u^r$.

For the proof of property (iii) of $V$ see \cite{Abramowiczetal95}.

Other possible choices of the ``radial velocity'' such as $u = |u^r|$ are not that
convenient.

\subsection{Description of accreting matter}

The stress-energy tensor $T^{ik}$ of the matter in the disk is given
by
\begin{equation}
 T^{ik} = \left(\varepsilon + p\right)u^i u^k
+ p\,g^{ik} + S^{ik} + u^k q^i + u^i q^k,
\label{tik}
\end{equation}
where $\varepsilon$ is the total energy density, $p$ is the
pressure,
\begin{equation}
S_{ik} =  \nu \rho \sigma_{ik}, \label{visc_tens}
\end{equation}
is the viscous stress tensor, $\rho$ is the rest mass density and
$q^i$ is the radiative energy flux.  In the last equation $\nu$ is
the kinematic viscosity coefficient and $\sigma_{ik}$ is the shear
tensor of the velocity field. From the first law of thermodynamic it
follows that
\begin{equation}
 d\varepsilon = {{\varepsilon + p}\over \rho}d\rho + \rho T dS,
\end{equation}
where $T$ is the temperature and $S$ is the entropy per unit mass.
Note, that in the physical units $\varepsilon = \rho c^2 +
\Pi$, where $\Pi$ is the internal energy. For non-relativistic
fluids, $\Pi \ll \rho c^2$ and $p \ll \rho c^2$, and therefore
\begin{equation}
 \varepsilon + p \approx c^2 \rho.
\end{equation}
We shall use this approximation (in geometrical units $\varepsilon +
p \approx \rho$) in all our calculations. This approximation does
not automatically ensure that the sound speed is below $c$, and one
should check this a posteriori when models are constructed. We write
the first law of thermodynamics in the form: \begin{equation}
       dU = -p\, d\left({1 \over \rho}\right) + TdS,
\label{1law} \end{equation} where $U= \Pi / \rho$.

\section{Slim disc equations in Kerr geometry}

General-Relativistic effects play an important role in the physics
of thin ($H/r \ll 1$) accretion discs close to the black hole but
they determine the properties of slim ($H/r \lesssim 1$) discs.
We will derive the slim-disc equation and before discussing their
properties we will say few words about thin discs.

It is convenient to write the final form of all the slim disk
equations at the equatorial plane, $z = 0$. Only these equations
which do not refer to the vertical structure could be derived
directly from the quantities at the equatorial plane with no further
approximations. All other equations are approximated --- either by
expansion in terms of the relative disk thickness $H/r$, or by
vertical averaging. 

\subsection{Mass conservation equation}

From general equation of mass conservation,
\begin{equation}
 \nabla^i \left (\rho u_i\right) = 0,
\end{equation}
and definition of the surface density $\Sigma$,
\begin{equation}
 \Sigma = \int^{+H(r)}_{-H(r)} \rho(r,z)dz \approx 2H \rho,
\label{Sigm}
\end{equation}
we derive the mass conservation equation,
\begin{equation}
 {\dot M} = - 2\pi\Delta^{1/2}\Sigma{V\over\sqrt{1-V^2}}.
\label{mass_c}
\end{equation}
  
In the Newtonian limit the mass conservation equation is:
\begin{equation}
{\dot M} = - 2\pi\Sigma\, v_r.
\end{equation}

\subsection{Equation of angular momentum conservation}

From the general form of the angular momentum conservation,
\begin{equation}
\nabla_k \left ( T^{ki}\xi_i \right) = 0,
\end{equation}
we derive, after {some} algebra,
\begin{equation}
{{\dot M}\over 2\pi r}{d{\mathfrak{L}}\over dr}+{1\over r}{d\over dr}
\left(\Sigma\nu A^{3/2} {\Delta^{1/2}\gamma^3 \over
r^4}{d\Omega\over dr}\right)-F^-{\mathfrak{L}}=0, \label{angm_c}
\end{equation}
where $F^-=2q_z$ is the vertical flux of radiation, and
\begin{equation}
{\mathfrak{L}}\equiv - (u \xi) = -u_{\varphi}=\gamma\left(A^{3/2}\over
r^3\Delta^{1/2}\right) {\tilde \Omega}, \label{spangm}
\end{equation}
is the {specific (per unit mass)} angular momentum.
The term $F^-{\cal L}$
represents angular momentum losses through radiation. Although it
was always fully recognized that angular momentum may be lost this
way, it has been argued that this term must be very small. Rejection
of this term enormously simplifies numerical calculations,
because with $F^-{\mathfrak{L}}=0$ equation
(\ref{angm_c}) can be trivially integrated,
\begin{equation}
 {{\dot M}\over {2\pi}}({\mathfrak{L}} - {\mathfrak{L}}_0) = -
\Sigma\nu A^{3/2} {\Delta^{1/2}\gamma^3\over r^4}{d\Omega\over dr}
\equiv \frac{\mathfrak{T}}{2\pi},
\label{amint}
\end{equation}
where ${\mathfrak{L}}_0$ is the specific angular momentum of matter \textsl{at the
horizon }($\Delta = 0$). In the  numerical scheme for integrating the
slim Kerr equations (with $F^-{\mathfrak{L}}$ assumed to be zero) the
quantity ${\mathfrak{L}}_0$ plays an important role: it is the eigenvalue
of the solutions that passes regularly through the sonic point.
The rhs of Eq. (\ref{amint}) $\mathfrak{T}$ represent the viscous torque transporting
angular momentum.

In the Newtonian case, a geometrically thin disc is Keplerian
$\Omega \approx\Omega_K$ (see Eq. (\ref{n_euler}), $\mathfrak{L}_K=\ell_K=R^2\Omega_K$,
Eq. (\ref{amint}) takes the familiar form of
\begin{equation*}
\nu \Sigma = \frac{\dot M}{3\pi}\left[1 - \frac{{\ell}_0}{\ell_K}\right],
\end{equation*}
(see Eq.\ref{eq:amintK}).

\subsection{Equation of momentum conservation}

From the r-component of the equation $\nabla_i T^{ik} = 0$ one
derives
\begin{equation}
{V\over1-V^2}{dV\over dr}={{\cal A} \over r} -{1\over\Sigma}{dP\over
dr},
\label{euler}
\end{equation}
where $P=2Hp$ is the vertically integrated pressure and
\begin{equation}
 {\cal A} = -{MA\over r^3\Delta\Omega_k^+\Omega_k^-}
{(\Omega-\Omega_k^+)(\Omega-\Omega_k^-)\over1-{\tilde
\Omega}^2{\tilde R}^2}.
\label{AA}
\end{equation}
Note that in Eq. (\ref{euler}) the viscous term has been neglected.

The Newtonian limit of Eq.(\ref{euler}) is
\begin{equation}
v_r\frac{d v_r}{dr} -\left(\Omega^2 - \Omega_K^2\right)r + \frac{c^2_s}{r}=0
\label{n_euler}
\end{equation}
For a thin disc: $H/r \approx c^2_s/r\Omega_K \ll 1$ Eq. (\ref{n_euler})
is simply $\Omega \approx \Omega_K$, i.e. a thin Newtonian disc
is Keplerian. The thickness of the disc depends on the efficiency of
radiative processes: efficient radiative cooling implies a low speed of
sound.

\subsection{Equation of energy conservation}

From the general form of the energy conservation
\begin{equation}
 \nabla_i \left( T^{ik}\eta_k\right ) = 0,
\end{equation}
and the first law of thermodynamics,
\begin{equation} T = {1\over \rho}
\left({{\partial\varepsilon}\over {\partial S}}\right)_{\rho},~~ p =
\rho \left({{\partial\varepsilon}\over {\partial \rho}}\right)_S -
\varepsilon,
\end{equation}
the energy equation can be written in general as
\begin{equation}
Q^{\rm adv}=Q^+ - Q^-, \label{en_c}
\end{equation}
where
\begin{equation}
Q^+=\nu\Sigma{A^2\over r^6}\gamma^4\left({d\Omega\over dr}\right)^2
\label{vish}
\end{equation}
is the surface viscous heat generation rate, $Q^-$ is the radiative cooling
flux (both surfaces) which is discussed in Section 5, and $Q^{\rm
adv}$ is the advective cooling rate due to the radial motion of the
gas. It is expressed as
\begin{equation}
{Q^{\rm adv}={\Sigma V\over\sqrt{1-V^2}}{\Delta^{1/2}\over
r}T{dS\over dr} \equiv{-{\dot{M} \over 2\pi r}T{dS\over dr}.}}
\label{adv}
\end{equation}

In \textsl{stationary} accretion flows advection is important only
in the inner regions close to the compact accretor. In the rest of the
flow the energy equation is just
\begin{equation}
Q^+=Q^-.
\label{en_th}
\end{equation}
  
In the newtonian limit of  Eqs. (\ref{vish}) and (\ref{amint})
one obtains
\begin{equation}
Q^+=\frac{3}{8\pi}\frac{GM\dot M}{R^3}\left(1 - \frac{\ell_0}{\ell}\right) .
\tag{\ref{eq:teff}}
\end{equation}

\subsection{Equation of vertical balance of forces}
\label{sub:verteq}

The equation of vertical balance is obtained by projecting the conservation equation
onto the $\theta$ direction
\begin{equation}
\label{eq: vert_project}
h^i_\theta\,\nabla_k T^k_i=0, \ \ \ \ \ \ \ \ \ \ \mathrm{where} \ \ \ \ \ \ \ \ \ \ \ h^i_\theta=\delta^i_\theta -u^iu_\theta
\end{equation}
and neglecting the terms $\mathcal{O}^3(\cos\theta)$.
For a non-relativistic fluid this leads to
\begin{equation}
\label{eq:vertrel}
\frac{dP}{dz}= - \rho g_z z= - \rho \frac{{\mathfrak L}^2 - a^2\left({\mathfrak E}^2 -1 \right)}{r^4}z\, .
\end{equation}
  
In the newtonian limit Eq. (\ref{eq:vertrel}) becomes
\begin{equation*}
\frac{dP}{dz}= - \rho \frac{{\ell_K}^2}{r^4}z  \ \ \ \ \ \ \ \ \ \ \ \ \ \ \ \   \ \ \ \ \ \ \ \ \ \ \ \ \ \ \ \ \ \ \ \ \mathrm{(see\,\,Eq. \,\ref{eq:vertacc})}.
\end{equation*}

\section{The sonic point and the boundary conditions}

\subsection{The ``no-torque condition"}

There have been a lot of discussion about the inner boundary condition
in an accretion disc. The usual reasoning is that for a thin disc the
inner boundary is at ISCO and since it is where circular orbits end
the boundary condition should be simply that the ``viscous" torque
vanishes (there is no orbit below the ISCO to interact with). 
Several authors have challenged this conclusion but a very simple argument by Bohdan Paczy\'nski \cite{Paczynski00}
shows the fallacy of these challenges.\\

Using Eq. (\ref{mass_c}), one obtains from Eq. (\ref{amint})
\begin{equation}
v^{(r)}=\nu \frac{A^{3/2}\gamma^2}{r^4}\frac{1}{{\mathfrak L} - {\mathfrak L}_0}.
{d\Omega\over dr}
\label{vr1}
\end{equation}

Next, from the viscosity prescription  $\nu \approx \alpha H^2 \Omega$,
and taking for simplicity the non-relativistic approximation (this does not
affect the validity of the argument but allows skipping irrelevant in this context
multiplicative factors) one can write
\begin{equation}
v_r \approx \alpha ~ H^2 ~ \frac{{\ell}}{{\ell} - {\ell}_0} ~ {d \Omega \over dr }
\approx \alpha ~ H^2 ~ \frac{{\ell}}{{\ell} - {\ell}_0} ~ { \Omega \over r }
\approx \alpha ~ v_{\varphi} \left( { H \over R} \right) ^2 ~ \frac{{\ell}}{{\ell} - {\ell}_0},
\label{vr2}
\end{equation}
where $v_{\varphi}= R \Omega$
Although we have dropped the GR terms, the  equation (\ref{vr2}) does not assume that the
radial velocity is small, i.e.this equation holds within the disk as well as
within the stream below the ISCO.  

Far out in the disk, where $ {\ell} \gg {\ell}_0$, one
obtains the standard formula (see Eq. \ref{eq:vvisc})
\begin{equation}
v_r \approx \alpha ~ v_{\varphi} \left( { H \over R } \right) ^2 ,
\hskip 1.0cm R \gg R_{in} .
\end{equation}

The flow crosses the black hole surface at the speed of light and since it is subsonic in the
disc it must somewhere become transonic, i.e. to go through a sonic point, close to
disc's inner edge.

At the sonic point we have $ v_r = c_s \approx (H/R)v_{\varphi} $, and
the equation (\ref{vr2}) becomes:
\begin{equation}
{v_r \over c_s} = 1 \approx \alpha ~ {H_{\rm in} \over R_{\rm in} } ~
\frac{{\ell}_{\rm in}}{{\ell}_{\rm in} - {\ell}_0},
\hskip 1.0cm R = R_{\rm in}
\label{bep1}
\end{equation}
If the disk is thin, i.e. $ H_{\rm in} / R_{\rm in} \ll 1 $, and the
viscosity is small, i.e.  $ \alpha \ll 1 $, then Eq. (\ref{bep1}) implies that
$({\ell}_{\rm in} - {\ell}_0)/{{\ell}_{\rm in}} \ll 1 $,
i.e. the specific angular momentum at the sonic point is almost
equal to the asymptotic angular momentum at the horizon.

In a steady state disk the torque ${\mathfrak T}$ has to satisfy the equation
of angular momentum conservation (\ref{amint}), which can be written as
\begin{equation}
{\mathfrak T}= \dot M \left( {\ell} -  {\ell}_0 \right) ,
\hskip 1.0cm
{\mathfrak T}_{in} = \dot M
\left({\ell}_{in} -  {\ell}_0 \right) .
\label{torquebp}
\end{equation}
  
Thus it is clear that for a thin, low viscosity disk the `no torque inner
boundary condition' (${\mathfrak T}_{in}\approx 0$) is an excellent approximation
\textit{following from angular momentum conservation}.

However, if the
disk and the stream are thick, i.e.  $ H/r \sim 1 $, and the
viscosity is high, i.e. $ \alpha \sim 1 $, then the angular momentum
varies also in the stream in accordance with the simple reasoning presented
above. However, the no--stress condition  at the disc inner edge might be
not satisfied.

\textsc{Additional reading}: Reference \cite{AfshordiPaczynski03}.
\\

\textsc{Acknowledgements}
I am grateful to Cosimo Bambi for having invited me to teach at the 2014 Fudan Winter School in Shanghai.
Discussions with and advice of Marek Abramowicz, Tal Alexander, Omer Blaes and Olek S{\c a}dowski were of great help. I thank
the Nella and Leon Benoziyo Center for Astrophysics at the Weizmann Institute for its hospitality in December 2014/January 2015 when parts of these lectures
were written. This work has been supported in part by the French Space Agency CNES and by the National Science Centre, Poland grants DEC-2012/04/A/ST9/00083,
UMO-2013/08/A/ST9/00795 and  UMO-2015/19/B/ST9/01099.

\section*{Appendix}
\addcontentsline{toc}{section}{Appendix}
\label{append}
{\sl Thermodynamical relations} \\

The equation of state can be expressed in the form:
\begin{equation}
P = P_r + {{\cal R} \over \mu_i } \rho T_i + {{\cal R} \over
\mu_e } \rho T_e + {B^2 \over 24 \pi},
\label{eq:eqst}
\end{equation}
where $P_r$
is the radiation pressure, $\cal R$ is the gas constant,
$\mu_i$ and $\mu_e$ the mean molecular weights of ions and electrons
respectively, $T_i$, and $T_e$
ion and electron temperatures, $a$ the radiation constant (not to be confused with
the dimensionless angular momentum $a$ in the Kerr metric), and $B$
the intensity of a isotropically tangled magnetic field, includes
the radiation, gas and magnetic pressures. The radiation pressure
$P_r$, the gas pressure $P_g$, and the magnetic pressure $P_m$
correspond respectively to the first term, the second and third
terms, and the last term in equation ($\ref{eq:eqst}$).

The mean molecular weights of ions and electrons can be well
approximated by:
\begin{equation} \mu_i \approx {4 \over 4X + Y}, \ \ \ \ \ \ \ \
\mu_e \approx {2 \over 1+X},
\label{eq:mus}
\end{equation}
where $X$ is the
relative mass abundance of hydrogen and $Y$ that of helium. We may
define a temperature as
\begin{equation}
T=\mu \left({T_i \over \mu_i} +{T_e
\over \mu_e}\right), \label{TTT} \end{equation} where \begin{equation} \mu = \left( {1\over
\mu_i} + {1\over \mu_e}\right)^{-1} \approx {2 \over 1 + 3X + 1/2Y}
\end{equation}
is the mean molecular weight. In the case of a one-temperature gas ($T_i=T_e$),
one has $T=T_i=T_e$. For an optically thick gas, $P_r=({4\sigma}/{3c}) 
T_r^4$.

For the
frozen-in magnetic field pressure $P_m \sim B^2 \sim \rho^{4/3}$,
therefore we may write the internal energy as
\begin{equation} U =  \frac{4\sigma }
{\rho c} T_r^4 + {{\cal R} T \over \mu m_u (\gamma_g - 1) }  + e_o
\rho^{1/3},
\label{eq:U}
\end{equation}
where $e_o$ is a constant ($P_m = 1/3 e_o
\rho^{4/3}$) and $\gamma_g$ is the ratio of the specific heats of
the gas. We define \begin{equation} \beta = {P_g \over p}, ~~\beta_m = {P_g \over
P_g + P_m}, ~~ \beta^*={4-\beta_m \over 3\beta_m} \beta. \end{equation} From
equations (\ref{eq:eqst}) and (\ref{eq:U}) one obtains the following
formulae (see e.g. ``Cox \& Giuli" 2004) for the specific heat at
constant volume:
\begin{equation}
c_V =  {{\cal R} \over \mu (\gamma_g - 1)}
       \left[{12 (1 - \beta/\beta_m)(\gamma_g - 1)+\beta \over
       \beta}\right]=
       {4 - 3\beta^* \over \Gamma_3 - 1} {P \over \rho T}
\end{equation}
and the adiabatic indices:
\begin{equation}
\Gamma_3 - 1 = {(4 -
3\beta^*)(\gamma_g - 1)
               \over 12(1 - \beta/\beta_m)(\gamma_g -1) + \beta}
\end{equation}

\begin{equation}
\Gamma_1 = \beta^* +(4 - 3\beta^*)(\Gamma_3 - 1).
\end{equation}

The ratio
of specific heats is $\gamma=c_p/c_V= \Gamma_1/\beta$. For
$\beta=\beta_m$ we have $\Gamma_3=\gamma_g$ and $\Gamma_1=(4 -
\beta)/3 + \beta(\gamma_g -1)$. For an equipartition magnetic field
($\beta=0.5$) one gets $\Gamma_1 = 1.5$ and for $\beta=0.95$,
$\Gamma_1= 1.65$ (here we have used $\gamma_g=5/3$).
One expects $\beta_m \sim 0.5 - 1$.
Since
\begin{equation}
T{dS \over dR}=c_V \left[{d \ln T\over dR} - (\Gamma_3 -1)
\left({d \ln \Sigma \over dR} - {d \ln H\over dR }\right)\right],
\end{equation}
The advective flux is written in the form:
\begin{equation}
Q^{\rm adv} = {\Mdot \over 2\pi R^2} {P\over \rho} \xi_a
\label{fadv}
\end{equation}
where
\begin{equation}
\xi_a = -\left[{4 - 3\beta^* \over \Gamma_3 - 1}
    {d \ln T\over d\ln R} + (4 - 3\beta^*)
     {d \ln \Sigma \over d \ln R}\right].
     \label{xi_a}
\end{equation}
The term $\propto d \ln H/d \ln R$ has been neglected. Since no
rigorous vertical averaging procedure exists, the presence or not of
the $d \ln H/d \ln R$ -- type terms in this (and other) equation may
be decided only by comparison with 2D calculations.

The formulae derived in this section are valid for the
optically thin case $\tau=0$ if one assumes $\beta=\beta_m$.
\vskip 0.5truecm
\textsc{Reference}: Weiss, A., Hillebrandt, 
W., Thomas, H.-C., 
\& Ritter, H.\ 2004, \textit{Cox and Giuli's  Principles of  Stellar Structure}, Cambridge, UK: Princeton Publishing Associates Ltd, 2004.

\end{document}